\documentclass[twocolumn]{aa}
\usepackage{graphicx}
\usepackage{txfonts}
\begin{document}
   \title{Is the dark matter halo of the Milky Way flattened?}

   \author{A. R\accent23u\v{z}i\v{c}ka
          \inst{1,}\inst{2},
          J. Palou\v{s}
          \inst{1}
          ,\and
          C. Theis
          \inst{3}
          }

   \offprints{A. R\accent23u\v{z}i\v{c}ka}
          
          \institute{Astronomical Institute, Academy of Sciences of the Czech Republic, 
                 Bo\v{c}n\'i II 1401a, 141 31 Prague, Czech Republic\\
            \email{adam.ruzicka@gmail.com}\\
            \email{palous@ig.cas.cz}
	    \and Faculty of Mathematics and Physics of the Charles
            University, Ke Karlovu 3, 121 16 Prague,
            Czech Republic
	    \and Institut f\"{u}r Astronomie der
            Universit\"{a}t Wien,
            T\"urkenschanzstrasse 17, A-1180 Wien, Austria\\
            \email{theis@astro.univie.ac.at}
             }

   \date{Received May 4, 2006; accepted July 25, 2006}

   \abstract{We performed an extended analysis of the parameter space for the interaction of the Magellanic
System with the Milky Way (MW). The varied parameters cover the phase space parameters,
the masses, the structure, and the orientation of both Magellanic Clouds, as well as the
flattening of the dark matter halo of the MW. The analysis was done by
a specially adopted optimization code searching for the best match between
numerical models and the detailed H\,I map of the Magellanic System by Br\"uns et al.~(\cite{Bruens05}).
The applied search algorithm is a genetic algorithm combined with a code based on
the fast, but approximative restricted N--body method. By this, we were able to 
analyze more than $\mathrm{10^6}$ models, which makes this study one of the most extended ones
for the Magellanic System. Here we focus on the flattening $q$ of the axially symmetric
MW dark matter halo potential, that is studied within the range $0.74 \leq q \leq 1.20$.
We show that creation of a trailing tail (Magellanic Stream) and a leading stream (Leading Arm) is
quite a common feature of the Magellanic System--MW interaction, and such structures were modeled across the entire
range of halo flattening values. However, important differences exist between the models, concerning
density distribution and kinematics of H\,I, and also the dynamical evolution of the Magellanic System.
Detailed analysis of the overall agreement between modeled and observed
distribution of neutral hydrogen shows that the models assuming an oblate ($q < 1.0$) dark matter halo
of the Galaxy allow for better satisfaction of H\,I observations than models with other halo configurations.
   \keywords{methods: N--body simulations -- Galaxy: halo -- galaxies: interactions -- galaxies: Magellanic Clouds}
   }

   \maketitle
%

\section{Introduction}
The idea of \emph{dark matter} (DM) was introduced by
Zwicky~(\cite{Zwicky33}). His dynamical measurements of the
mass--to--light ratio of the Coma cluster gave larger values than
those known from luminous parts of nearby spirals. That discrepancy
was explained by the presence of DM. Ostriker et
al.~(\cite{Ostriker74}) proposed that DM is concentrated in a form of
extended galactic halos. Analysis of rotation curves of spiral
galaxies (Bosma~\cite{Bosma81}; Rubin\,\&\,Burstein~\cite{Rubin85})
denotes that their profiles cannot be explained without presence of
non--radiating DM. Hot X--ray emitting halos have been used to
estimate total galactic masses (McLaughlin~\cite{McLaughlin99}).
Corresponding mass--to--light ratios exceed the maximum values
for stellar populations, and DM explains the missing matter naturally.
The presence of DM halos is expected by the standard CDM cosmology model
of hierarchical galaxy formation.  The classical CDM halo profile
(NFW) is simplified to be spherical. However, most CDM models expect
even significant deviations from the spherical symmetry of DM distribution
in halos.  The model of formation of DM halos in the universe
dominated by CDM by Frenk et al.~(\cite{Frenk88}) produced triaxial
halos with a preference for prolate configurations.
Numerical simulations of DM halo formation by
Dubinski\,\&\,Carlberg~(\cite{Dubinski91}) are consistent with halos
that are triaxial and flat, with (c/a) = 0.50 and (b/a)
= 0.71. There are roughly equal numbers of dark halos with oblate and
prolate forms. 

Observationally, the measurement of the shape of a DM halo is a
difficult task. A large number of varying techniques found
notably different values, and it is even not clear if the halo
is prolate or oblate. Olling\,\&\,Merrifield~(\cite{Olling00})
use two approaches to investigate the DM halo
shape of the Milky Way (MW), a rotation curve analysis and the radial
dependence of the thickness of the H\,I layer. Both methods
lead consistently to flattened oblate halos.

Recently, the nearly planar distribution of the observed MW
satellites, which is almost orthogonal to the Galactic plane, raised
the question if they are in agreement with cosmological CDM models
(Kroupa et al.~\cite{Kroupa05}) or if other origins 
have to be invoked.  Zentner et al.~(\cite{Zentner05})
claim that the disk--like distribution of the MW satellites can be
explained, provided the halo of the MW is sufficiently prolate
in agreement with their CDM simulations.
On the other hand, it is not clear if there exists a unique prediction 
of the axis ratios from CDM simulations, as the scatter in 
axis ratios demonstrates (Dubinski\,\&\,Carlberg~\cite{Dubinski91}).
Based on $\Lambda$CDM simulations, Kazantzidis et al.~(\cite{Kazantzidis04}) 
emphasize that gas cooling strongly affects halo shapes with the 
tendency to produce rounder halos.

Another promising method to determine the Galactic halo shape are
stellar streams because they are coherent structures covering large
areas in space. Thus, their shape and kinematics should be strongly
influenced by the overall properties of the underlying potential.  A
good candidate for such an analysis is the stellar stream associated
with the Sagittarius dwarf galaxy. By comparison with simulations, Ibata
et al.~(\cite{Ibata01}) found that the DM halo is almost spherical in
the galactocentric distance range from 16 to 60 kpc. Helmi~(\cite{Helmi04})
studied the Sagittarius dwarf stream and argues for a non--spherical shape of
the Galactic halo. However, Helmi~(\cite{Helmi04}) also
warned that the Sagittarius stream might be
dynamically too young to allow for constraints on the halo shape.
Johnston et al.~(\cite{Johnston05}) argue that the analysis
of the orbital planes of the Sagittarius dwarf galaxy prefers oblate
over prolate models. In flattened oblate halos, the non--polar satellite orbits
tend to become co--planar and contribute to the disk formation. In the
case of Monoceros tidal stream discussed by Pe\~narrubia et al.~(\cite{Pena05}),
the best--fitting models predict a significantly oblate halo.

In this paper we use the Magellanic Stream to derive
constraints on the halo shape of the MW. The basis are the new
detailed H\,I observations of the Magellanic System (including the
Large Magellanic Cloud (LMC) and the Small Magellanic Cloud (SMC)) by
Br\"uns et al.~(\cite{Bruens05}).  As remnants of the LMC--SMC--MW
interaction, extended structures connected to the System are observed.
Among them, the \textit{Magellanic Stream} -- an H\,I tail
originating in between the Clouds and spreading over $\mathrm{\approx
100^\circ}$ of the plane of sky -- has been a subject of
investigation for previous studies (see, e.g.,\ 
Fujimoto\,\&\,Sofue~\cite{Sofue76};
Lin\,\&\,Lynden--Bell~\cite{Lin77};
Murai\,\&\,Fujimoto~\cite{Murai80};
Heller\,\&\,Rohlfs~\cite{Heller94}; 
Gardiner\,\&\,Noguchi~\cite{Gardiner96};
Bekki\,\&\,Chiba~\cite{Bekki05};
Mastropietro et al.~\cite{Mastropietro05}).  
Due to the
extended parameter space related to the interaction of three galaxies
and also due to the high computational costs of fully self--consistent
simulations, simplifying assumptions were unavoidable. Many simulations
neglected the self--gravity of the individual stellar systems by
applying a restricted N--body method similar to the method introduced by
Toomre\,\&\,Toomre~(\cite{Toomre72}). None of these simulations
considered the self--gravity of all three galaxies. Often only one
galaxy is simulated, including its self--gravity by means of a live disk and
halo, whereas the other two galaxies are taken into account by rigid
potentials of high internal symmetry. That is, none of the simulations so
far adopted a live dark matter halo of the MW, but they applied
\mbox{(semi--)analytical} descriptions for the \textit{dynamical friction} between the
Magellanic Clouds and the MW. Also, a possible flattening of
the MW halo has not been considered.  Having the numerical
difficulties in mind, it is not surprising that a thorough
investigation of the complete parameter space was impossible.

Modeling observed interacting galaxies means dealing with an extended
high--dimensional space of initial conditions and parameters of the
interaction. Wahde~(\cite{Wahde98}) and Theis~(\cite{Theis99}) 
introduced a \emph{genetic algorithm} (GA) as a robust search 
method to constrain models of observed interacting galaxies. 
The GA optimization scheme selects models
according to their ability to match observations. Inspired by their
results, we employed a simple fast numerical model of the Magellanic
System combined with an implementation of a GA to perform the first
very extended search of the parameter space for the interaction
between LMC, SMC, and MW. Here we present our results about the MW DM
halo flattening values compatible with most detailed currently
available H\,I Magellanic survey (Br\"uns et al.~\cite{Bruens05}, see Figs.~\ref{2d_HI_map},
and~\ref{pic_obs_radvel}).


\section{Magellanic Clouds and MW interaction}
\subsection{Observations of the Magellanic System}
\begin{figure}[h]
\centering
\includegraphics[bb=0 0 390 320, angle=90, width = 8cm, clip]{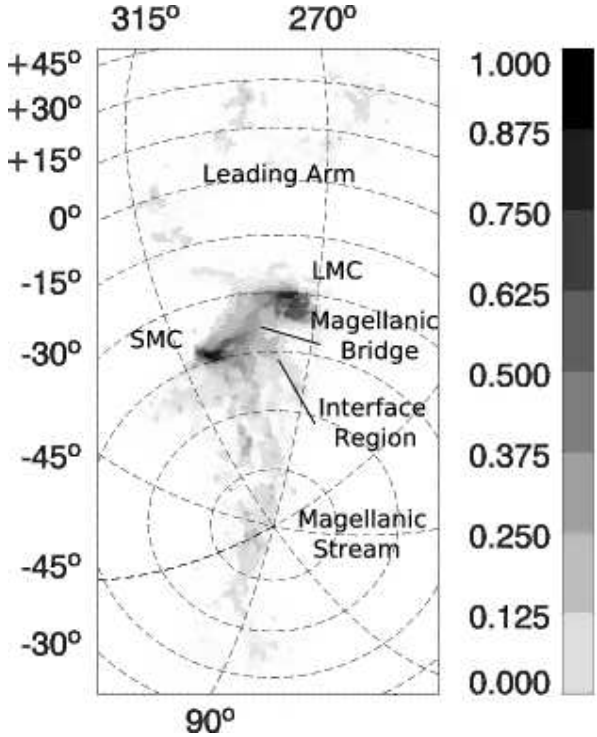}
\caption{Contour map of the observed
H\,I integrated relative column density in the Magellanic System.
Data by Br\"uns et al.~(\cite{Bruens05}) is projected on the plane of sky. Galactic coordinates are used.}
\label{2d_HI_map}
\end{figure}

\begin{figure}[h]
\centering
\includegraphics[bb=0 0 335 450, angle=90, width = 8cm, clip]{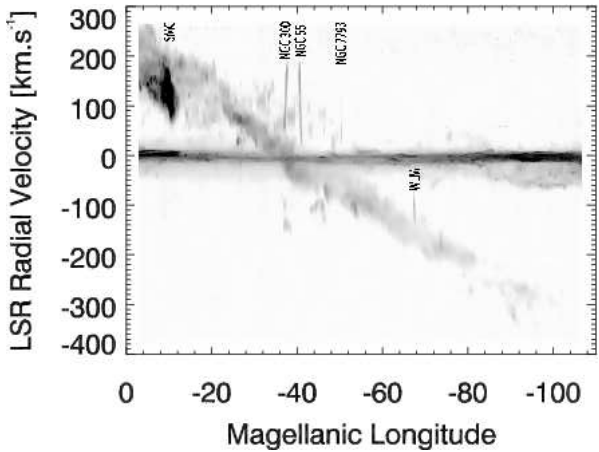}
\caption{LSR radial velocity of the Magellanic Stream as a function of Magellanic Longitude
(see Wannier\,\&\,Wrixon~\cite{WannierMS}). 
The observational data by Br\"{u}ns et al.~(\cite{Bruens05}) is plotted.
Strong H\,I emission observed for $\mathrm{\upsilon_{LSR}\approx 0\,km\,s^{-1}}$ comes 
from the MW. The map also shows the emission from the SMC, the galaxies NGC\,300, NGC\,55, and NGC\,7793 from the Sculptor Group, and the Local
Group galaxy WLM.}
\label{pic_obs_radvel}
\end{figure}

The MW, together with its close dwarf companions the LMC and SMC,
forms an interacting system.  Hindman et al.~(\cite{Hindman63})
observed the H\,I \emph{Magellanic Bridge} (MB) connecting the Clouds.
Another significant argument for the LMC--SMC--MW interaction was
brought by Wannier\,\&\,Wrixon~(\cite{WannierMS}) and Wannier et
al.~(\cite{WannierLA}). Their H\,I observations of the Magellanic
Clouds discovered large filamentary structures projected on the plane
of the sky close to the Clouds, and extended to both high negative and
positive radial LSR velocities.  Mathewson et al.~(\cite{Mathewson74})
detected another H\,I structure and identified a narrow tail emanating
from the space between the LMC and SMC, spread over the South Galactic
Pole. The tail was named the Magellanic~Stream. A similar H\,I
structure called the \emph{Leading~Arm} extends to the north of
the Clouds, crossing the Galactic plane. A high--resolution, spatially
complete H\,I survey of the entire Magellanic System done by Br\"{u}ns
et al.~(\cite{Bruens05}, see Figs.~\ref{2d_HI_map}, and~\ref{pic_obs_radvel})
gives detailed kinematic
information about
the Clouds and the connected extended structures. It indicates that
the observed features consist of the matter torn off the Magellanic
Clouds and spread out due to the interaction between the LMC, SMC, and
MW.

\subsection{Modeling of the Magellanic System}\label{models}
Toomre\,\&\,Toomre~(\cite{Toomre72}) have shown the applicability of
\emph{restricted N--body} models on interacting galaxies.  In
restricted N--body simulations, gravitating particles are replaced by
\emph{test--particles} moving in a time--dependent potential that is
a superposition of analytic potentials of the individual galaxies.
Such an approach maps the gravitational potential with high spatial
resolution for low CPU costs due to the linear CPU scaling with the
number of particles. However, the self--gravity of the stellar systems
is not considered directly. That is, the orbital decay of the 
Magellanic Clouds due to dynamical friction cannot be treated 
self--consistently in restricted N--body simulations, but has to be
considered by (semi--)analytical approximative formulas.

The first papers on the physical features of the interacting system of
the LMC, the SMC and the Galaxy used 3\,D restricted N--body simulations to
investigate the \textit{tidal origin} of the extended Magellanic
structures.  Lin\,\&\,Lynden--Bell~(\cite{Lin77}) pointed out the
problem of the large parameter space of the LMC--SMC--MW interaction.
To reduce the parameter space, they neglected both the SMC influence
on the System and dynamical friction within the MW halo, and showed that such
configuration allows for the existence of a LMC trailing tidal stream.
The interaction between the Clouds was analyzed by
Fujimoto\,\&\,Sofue~(\cite{Sofue76}), who assume the LMC and SMC to
form a gravitationally bound pair for several Gyr, moving in a
flattened MW halo. They identified some LMC and SMC orbital
paths leading to the creation of a tidal tail.  Following studies by
Murai\,\&\,Fujimoto~(\cite{Murai80}, \cite{Murai84}),
Lin\,\&\,Lynden--Bell~(\cite{Lin82}), Gardiner et
al.~(\cite{Gardiner94}), and Lin et al.~(\cite{Lin95}) extended and
developed test--particle models of the LMC--SMC--MW interaction. The
Magellanic Stream was reproduced as a remnant of the LMC--SMC encounter that was
mostly placed at the time of $\mathrm{-2\,Gyr}$. The matter torn off
was spread along the paths of the Clouds. The simulations also
indicate that the major fraction of the Magellanic Stream gas stems from the SMC.
The observed radial velocity profile of the Stream was modeled
remarkably well. However, the smooth H\,I column density distribution
did not agree with observations indicating apparently clumpy Magellanic Stream
structure.
Test--particle models place matter
at the Leading Arm area naturally (see the study on the origin of tidal tails and arms by Toomre\,\&\,Toomre~(\cite{Toomre72})),
but correspondence with observational data
cannot be considered sufficient.
Gardiner\,\&\,Noguchi~(\cite{Gardiner96}) devised a scheme of the
Magellanic System interaction implementing the full N--body approach.
The SMC was modeled by a self--gravitating sphere moving in the LMC and MW
analytic potentials. It was shown that the evolution of the Magellanic Stream and the Leading Arm is
dominated by tides, which supports the applicability of test--particle
codes for the modeling of extended Magellanic structures.
Recently, the study by Connors et al.~(\cite{Connors05}) investigated the evolution of the Magellanic Stream as
a process of tidal stripping of gas from the SMC. Their high--resolution N--body model of the Magellanic System
based on ideas of Gardiner\,\&\,Noguchi~(\cite{Gardiner96}) is compared to the data from the HIPASS survey.
Involving pure gravitational interaction allowed for remarkably good reproduction of the Magellanic Stream
LSR radial velocity profile. They were able to improve previous models of the Leading Arm. Similarly to the
previous tidal scenarios, difficulties remain concerning overestimations of the H\,I column density toward the far tip
of the Magellanic Stream. Connors et al.~(\cite{Connors05}) approximate both the LMC and the MW by rigid potentials
and also do not study the influence of the non--spherical halo of the MW.

Meurer et al.~(\cite{Meurer85}) involved continuous ram pressure
stripping into their simulation of the Magellanic System. This
approach was followed later by Sofue~(\cite{Sofue94}), who neglected
the presence of the SMC, however. The Magellanic Stream was formed of the gas stripped from
outer regions of the Clouds due to collisions with the MW extended
ionized disk. Heller\,\&\,Rohlfs~(\cite{Heller94}) argue for a
LMC--SMC collision 500\,Myr ago that established the MB. Later, gas
distributed to the inter--cloud region was stripped off by ram
pressure as the Clouds moved through a hot MW halo. Generally,
continuous ram pressure stripping models succeeded in reproducing the
decrease of the Magellanic Stream H\,I column density towards the far tip of the
Stream. However, they are unable to explain the evidence of gaseous
clumps in the Magellanic Stream.
Gas stripping from the Clouds caused by isolated collisions in
the MW halo was studied by Mathewson et al.~(\cite{Mathewson84}). The
resulting gaseous trailing tail consisted of fragments, but such a
method did not allow for reproduction of the column density decrease
along the Magellanic Stream.  Recently, Bekki\,\&\,Chiba~(\cite{Bekki05}) applied a
complex gas--dynamical model including star--formation to investigate
the dynamical and chemical evolution of the LMC.  They include
self--gravity and gas dynamics by means of sticky particles, but they
are also not complete: they assume a live LMC system, but the SMC and MW
were treated by static spherical potentials. Thus, dynamical friction of the LMC in
the MW halo is only considered by an analytical formula and a possible
flattening of the MW halo is not involved. Their model cannot
investigate the possible SMC origin of the Magellanic Stream either.  Mastropietro et
al.~(\cite{Mastropietro05}) introduced their model of the Magellanic
System including hydrodynamics (SPH) and a full N--body description of
gravity. They studied the interaction between the LMC and the MW.
Similarly to Lin\,\&\,Lynden--Bell~(\cite{Lin77}) and
Sofue~(\cite{Sofue94}), the presence of the SMC was not taken into account. It
was shown that the Stream, which sufficiently reproduces the observed H\,I
column density distribution, might have been created without an
LMC--SMC interaction. However, the history of the Leading Arm was not
investigated.

In general, hydrodynamical models allow for better reproduction of the Magellanic Stream H\,I column density profile than tidal schemes.
However, they constantly fail to reproduce the Magellanic Stream radial velocity measurements and especially the high--negative
velocity tip of the Magellanic Stream. Both families of models suffer from serious difficulties when modeling the Leading Arm.

To model the evolution of the Magellanic System, the initial
conditions and all parameters of their interaction have to be
determined. Such a parameter space becomes quite extended.
In the Magellanic System we have to deal with the orbital
parameters and the orientation of the two Clouds, their internal properties
(like the extension of the disk), and the properties of the MW potential.
In total we have about 20
parameters (the exact number depends on the adopted sophistication of the
model).
Previous studies of the Magellanic Clouds, however, argued for
very similar evolutionary scenarios of the system (e.g.,
Lin\,\&\,Lynden--Bell~\cite{Lin82}, Gardiner et al.~\cite{Gardiner94}, 
Bekki\,\&\,Chiba~\cite{Bekki05}). These calculations are based on
additional assumptions concerning the
orbits, or the internal structure and orientation of the Clouds, the
potential of the MW (mass distribution and shape), the
treatment of dynamical friction in the Galactic halo, or the treatment
of self--gravity and gas dynamics in the Magellanic Stream. Some of them can be
motivated by additional constraints. For example, Lin\,\&\,Lynden-Bell~(\cite{Lin82}) 
and Irwin et al.~(\cite{Irwin90}) argue that the Clouds should have been
gravitationally bound over the last several Gyr.
However, in general, the uniqueness of the adopted models is unclear,
because a systematic analysis of the entire parameter space is 
still missing. 

We explore the LMC--SMC--MW interaction parameter space that is
compatible with the observations of the Magellanic System available
to date.  Regarding the dimension and size of the parameter space, a
large number of the numerical model runs have to be performed to test
possible parameter combinations, no matter what kind of search
technique is selected. In such a case, despite their physical
reliability, full N--body models are of little use, due to their
computational costs.


\section{Method}
To cope with the extended parameter space, neither a
complete catalog of models nor a large number of computationally expensive 
self--consistent simulations can be performed, both due to numerical
costs. However, a new numerical approach based on evolutionary optimization
methods combined with fast (approximative) N--body integrators turned out to
be a promising tool for such a task.  
In case of encounters between two galaxies, Wahde~(\cite{Wahde98}) and 
Theis~(\cite{Theis99}) showed that a combination
of a genetic algorithm with restricted N--body simulations is able to 
reproduce the parameters of the interaction.

Here we apply the GA search strategy with a restricted N--body code
for the Magellanic System. In the following sections we describe first
our N--body calculations and then we briefly explain the applied 
genetic algorithm.


\subsection{N--body simulations}
Our simulations were performed by test--particle codes 
similar to the ones already applied to the Magellanic System by
Murai\,\&\,Fujimoto~(\cite{Murai80}) and Gardiner et
al.~(\cite{Gardiner94}). But as an extension of these previous
papers we allow for a flattened MW halo potential and 
a more exact formula for dynamical friction taking anisotropic
velocity distributions into account.

For the galactic potentials we used the following descriptions:
both LMC and SMC are represented by Plummer spheres.
The potential of the DM halo of the MW is modeled
by a flattened axisymmetric logarithmic potential
(Binney\,\&\,Tremaine~\cite{Binney87})
\begin{equation}
  \Phi_\mathrm{L} = \frac{1}{2}{\upsilon_0}^2\ln\left({R_\mathrm{c}}^2 + R^2 + \frac{z^2}{q^2}\right).
  \label{log_gravity}
\end{equation}
In agreement with Helmi~(\cite{Helmi04})
we set $R_\mathrm{c} = 12\,\mathrm{kpc}$ and $\upsilon_0 = \sqrt{2} \cdot
131.5\,\mathrm{km\,s^{-1}}$, and $q$ describes the flattening of the
MW halo potential. The corresponding flattening $q_\rho$ of the density 
distribution associated with the halo potential follows
\begin{equation}
  {q_{\rho}}^2 = \frac{1 + 4q^2}{2 + 3/q^2}\ \ \ (R \ll R_\mathrm{c}),
  \label{log_dens_flattening_1}
\end{equation}
\begin{equation}
  {q_{\rho}}^2 = {q}^2\left(2 - \frac{1}{{q}^2}\right)\ \ \ (R \gg R_\mathrm{c}).
  \label{log_dens_flattening_2}
\end{equation}
The choice for the logarithmic potential was motivated by several reasons.
First, it allows for the investigation of flattened configurations of the Galactic halo
at low computational costs. Also the relatively small number of input parameters of Eq.~\ref{log_gravity}
is an advantage concerning the performance and speed of the numerical model. Finally, the logarithmic potential
was employed in the recent study by Helmi~(\cite{Helmi04}) that used the similar method of dwarf galaxy streams
to investigate the MW DM halo, and so the application of the same formula
allows for comparison of our results.

Dynamical friction causes the orbital decay of the Magellanic Clouds.  We adopted the
analytic dynamical friction formula by Binney~(\cite{Binney77}). In contrast to the
commonly used expression by Chandrasekhar~(\cite{Chandra43}), it
allows for an anisotropic velocity distribution. By comparison with
N--body simulations of sinking satellites, Pe\~{n}arrubia et
al.~(\cite{Pena04}) showed that Binney's solution is a significant
improvement over the standard approach with Chandrasekhar's formula.

Finally, we get the following equations of motion of the Clouds:
\begin{equation}
\frac{\mathrm{d}\vec{\upsilon}_\mathrm{LMC}}{\mathrm{d}t} = -(\nabla\Phi_\mathrm{L} + \nabla\Phi_\mathrm{SMC}) + \frac{\vec{F}_\mathrm{DF}}{m_\mathrm{LMC}},
\end{equation}
\begin{equation}
\frac{\mathrm{d}\vec{\upsilon}_\mathrm{SMC}}{\mathrm{d}t} = -(\nabla\Phi_\mathrm{L} + \nabla\Phi_\mathrm{LMC}) + \frac{\vec{F}_\mathrm{DF}}{m_\mathrm{SMC}},
\end{equation}
where $\Phi_\mathrm{LMC}$, $\Phi_\mathrm{SMC}$ are the LMC, SMC Plummer potentials, and $\vec{F}_\mathrm{DF}$ is the
dynamical friction force exerted on the Clouds as they move in the MW DM halo.

Our simulations were performed with the total number of 10\,000 test--particles equally distributed 
to both Clouds.
We start the simulation with test--particles in a disk--like 
configuration with an exponential particle density profile,
and compute the evolution of the test--particle distribution up to
the present time.
Initial conditions for the starting point
of the evolution of the System at the time --4\,Gyr were obtained by the standard backward integration of equations of motion (see, e.g.,
Murai\,\&\, Fujimoto~\cite{Murai80},
Gardiner et al.~\cite{Gardiner94}). 
Basically, the choice for the starting epoch of this study originates in the fact that the MW, LMC, and SMC form an isolated
system in our model. Such an assumption was very common in previous papers on the Magellanic System (e.g., Murai\,\&\,Fujimoto~\cite{Murai80},
Gardiner et al.~\cite{Gardiner94}, Gardiner\,\&\,Noguchi~\cite{Gardiner96}) and was motivated by insufficient kinematic information
about the Local Group, that did not allow for the estimation of the influence of its members other than MW on the evolution of the Magellanic System.
Our detailed analysis of the orbits of LMC and SMC showed that the galactocentric distance of either of the Clouds
did not exceed 300\,kpc within the last 4\,Gyr. Investigation of the kinematic history of the Local Group by Byrd et al.~(\cite{Byrd94})
indicates that the restriction of the maximal galactocentric radius of the Magellanic System to $R_\mathrm{max}\approx 300$\,kpc when $T>-4$\,Gyr
lets us consider the orbital motion of the Clouds to be gravitationally dominated by the MW.

\subsection{Genetic algorithm search}
Genetic algorithms can be used to solve optimization problems like a
search in an extended parameter space. The basic concept of GA
optimization is to interpret the natural evolution of a population of
individuals as an optimization process, i.e., an increasing adaptation
of a population to given conditions. In our case the conditions are
to match numerical models to the observations.  Each single point
in parameter space uniquely defines one interaction scenario that can
be compared with the observations (after the N--body simulation is
performed). The quality of each point in parameter space (or the
corresponding N--body model) can be characterized by the value of a {\it fitness
function} ($FF$), which is constructed to become larger, the better the
model matches the observations. A {\it population} consists of a
set of points in parameter space ({\it individuals}). Each single
parameter of an individual corresponds to a {\it gene}. 
A genetic algorithm recombines the genes of the individuals in
different reproduction steps: First two individuals ({\it parents}) 
are selected with a probability growing with their fitness. Then the
genes of the parents are recombined by application of reproduction 
operators mimicking {\it cross--over} and {\it mutation}. Often
the reproduction is done for all members of a population. Then
the newly created population corresponds to a next {\it generation}.
The reproduction steps are then repeated until a given number
\begin{figure}[h]
\centering
\includegraphics[bb=5 265 535 700, angle=90, width = 9.5cm, clip]{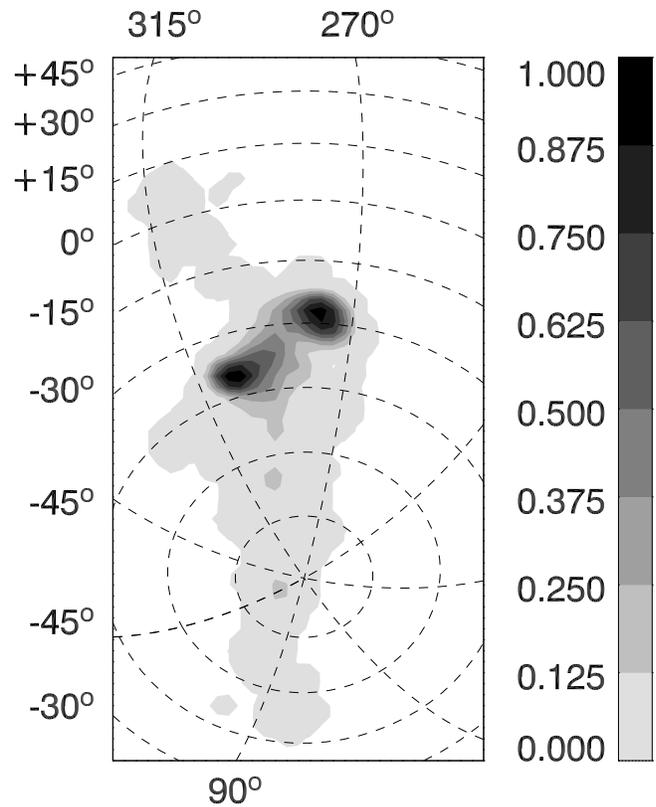}
\caption{Contour map of the observed H\,I integrated column density. The original
data cube by Br\"uns et al.~(\cite{Bruens05}) is modified by frequency filtering (see Appendix~\ref{appendixB}) and integrated
over all radial velocity channels. The contour map is projected on the plane of sky.
Galactic coordinates are used.}
\label{pic_obs_dens}
\end{figure}
of generations is calculated or a sufficient convergence
is reached. More details about genetic algorithms can be found in
Holland~(\cite{Holland75}) or Goldberg~(\cite{Goldberg89}). An application to interacting galaxies
is described in Wahde ~(\cite{Wahde98}), 
Theis~(\cite{Theis99}), or Theis\,\&\,Kohle~(\cite{Theis01}).

Obviously, the evolutionary search for the optimal solution can
be treated as a process of maximizing the fitness of individuals.
The GA looks for maxima of the function assigning individuals their 
fitness values. It should be noted that the problem we want to solve only
enters via the $FF$. Therefore, the choice of the
$FF$ is essential for the performance and the answer
given by a GA.

For our calculations we used a $FF$ consisting of
three different parts corresponding to three different comparisons.
These three $FF$s measure the quality of the numerical
models for different aspects of the given H\,I data cube. The original 3D H\,I data cube
together with its version modified for the purpose of an efficient GA search are visualized
in Fig.~\ref{3d_HI_map}. Two of the
comparisons deal with the whole data cube: $FF_2$ denotes the 
rough occupation of cells in the data cube and is a measure for
the agreement of the structural shape in the data cube. 
$FF_3$ compares the individual intensities in each cell of the
data cube. The $FF_3$ definition basically follows the $FF$ introduced
in Theis~(\cite{Theis99}) or Theis\,\&\,Kohle~(\cite{Theis01}), who found it to be an efficient
GA driver for the galactic interaction problems they studied. However, if the
fraction of the total
volume of the system's data cube that is occupied by the structures of special interest
is small ($\mathrm{<10\%}$
in the case of the Magellanic Stream and the Leading Arm), implementation of a structural search in the data
($FF_2$) preceding the fine comparison between modeled and observed data significantly improves the performance of the GA.
Finally, $FF_1$ is introduced to take into account the velocity profile,
i.e., the important constraint of the minimum velocity in the stream.
All these three quality measures are combined to 
yield the final fitness of a model. Details can be found in
Appendix~\ref{appendixB}.


\subsection{The GA parameter space}
In this paragraph we introduce the parameters of the Magellanic System
interaction that were subject to our GA search. The parameters involve the initial
conditions of the LMC and SMC motion, their total masses, parameters of mass distribution,
particle disc radii, and orientation angles, and also the MW dark matter halo potential flattening parameter.
Table~\ref{tab_parameters} summarizes the parameters of the interaction and introduces their current values.
Models are described in a right--handed Cartesian coordinate
\begin{table}[h]
\centering
\caption{Parameters of the MW--LMC--SMC interaction.}
\label{tab_parameters}
\begin{tabular}{lrl}
\hline
\hline                
\textbf{Param.} & \textbf{Value} & \\
\hline
$\mathrm{\vec{r}_{LMC}[kpc]}$ & $\left(\begin{array}{r}
			    	     \langle -1.5, -0.5 \rangle \\
				     \langle -41.0, -40.0 \rangle \\ 
				     \langle -27.1, -26.1 \rangle
				     \end{array}\right)$ & Current galactocentric\\
 & &				     				       position vectors\\
$\mathrm{\vec{r}_{SMC}[kpc]}$ & $\left(\begin{array}{r}
			    	     \langle 13.1, 14.1 \rangle \\
				     \langle -34.8, -33.8 \rangle \\
				     \langle -40.3, -39.3 \rangle
				     \end{array}\right)$ & \\
\hline
$\mathrm{\vec{\upsilon}_{LMC}[km\,s^{-1}]}$ & $\left(\begin{array}{r}
			    	     \langle -3, 85 \rangle \\
				     \langle -231, -169 \rangle \\ 
				     \langle 132, 206 \rangle
				     \end{array}\right)$ & Current velocity\\
 & & 				     					      vectors \\
$\mathrm{\vec{\upsilon}_{SMC}[km\,s^{-1}]}$ & $\left(\begin{array}{r}
			    	     \langle -112, 232 \rangle \\
				     \langle -346, -2 \rangle \\ 
				     \langle 45, 301 \rangle
				     \end{array}\right)$ & \\
\hline
$m_\mathrm{LMC}\mathrm{[10^9M_{\odot}]}$ & $\langle 15.0, 25.0\rangle$ & Masses\\
$m_\mathrm{SMC}\mathrm{[10^9M_{\odot}]}$ & $\langle 1.5, 2.5\rangle$ & \\
\hline
$\epsilon_\mathrm{LMC}\mathrm{[kpc]}$ & $\langle 2.5, 3.5 \rangle$ & Plummer sphere\\
$\epsilon_\mathrm{SMC}\mathrm{[kpc]}$ & $\langle 1.5, 2.5 \rangle$ & scale radii\\
\hline
$r^\mathrm{disk}_\mathrm{LMC}\mathrm{[kpc]}$ & $\langle 9.0, 11.0 \rangle$ & Particle disk radii\\
$r^\mathrm{disk}_\mathrm{SMC}\mathrm{[kpc]}$ & $\langle 5.0, 7.0 \rangle$ & \\
\hline
$\Theta^\mathrm{disk}_\mathrm{LMC}$ & $\langle 87^{\circ}, 107^{\circ} \rangle$ & Disk orientation angles \\
$\Phi^\mathrm{disk}_\mathrm{LMC}$ &  $\langle 261^{\circ}, 281^{\circ} \rangle$ & \\
$\Theta^\mathrm{disk}_\mathrm{SMC}$ & $\langle 35^{\circ}, 55^{\circ} \rangle$ & \\
$\Phi^\mathrm{disk}_\mathrm{SMC}$ & $\langle 220^{\circ}, 240^{\circ} \rangle$ & \\
\hline
$q$ & $\langle 0.74, 1.20\rangle$ & MW DM halo \\
 & & potential flattening\\
\hline
\end{tabular}
\end{table}
system with the origin in the Galactic center. This system is
considered to be an inertial frame because we assume that 
$m_\mathrm{MW} \gg  m_\mathrm{LMC}, m_\mathrm{SMC}$, where \(m_\mathrm{MW}\) is the mass of MW,
and \(m_\mathrm{LMC}\) and \(m_\mathrm{SMC}\) are the masses of LMC and SMC, respectively. Therefore, the center of
mass of the system may be placed at the Galactic center. We assume the present
position vector of the Sun $\mathrm{\vec{R_\odot} = (-8.5, 0.0, 0.0)\,kpc}$.
In the following paragraphs we will discuss the parameters of the LMC--SMC--MW interaction and the determination
of their ranges.

The nature and distribution
of dark matter in the Galaxy has been subject to intense research
and a large number of models have been proposed. 
We probe the DM matter distribution by investigating the redistribution of matter in the Magellanic System due to the MW--LMC--SMC
interaction, paying special attention to the features of the Magellanic Stream. That is similar to the method applied by
Helmi~(\cite{Helmi04}), who studied kinematic properties of the Sagittarius stream. To enable comparison with the results
by Helmi~(\cite{Helmi04}), we also adopted the axially symmetric logarithmic halo model of MW (Eq.~\ref{log_gravity}) and the
same values of the halo structural parameters $R_\mathrm{c}$, and $\upsilon_0$ with a similar range of studied values of the flattening $q$
(see Table~\ref{tab_parameters}).
We extended the range of $q$ values tested by Helmi~(\cite{Helmi04}) 
to the lower limit of $q=0.74$, which is the minimal value acceptable for the model of a logarithmic halo
(for a detailed explanation see Binney\,\&\,Tremaine~(\cite{Binney87})). 
For every value of $q$, a time--consuming calculation of parameters of dynamical friction is required (see Appendix~\ref{appendixA}).
To reduce the computational difficulties, the flattening $q$ was treated as a discrete value with a step of $\Delta q=0.02$, and
the parameters of dynamical friction were tabulated.
The upper limit of $q=1.20$ was selected to enable testing of prolate halo configurations. Extension of the halo flattening
range to higher values was not considered due to the computational performance limits of our numerical code.

The estimated
ranges of the values of the remaining parameters are based on our observational knowledge in the Magellanic System. Galactocentric position vectors $\vec{r}_\mathrm{LMC}$ and 
$\vec{r}_\mathrm{SMC}$ agree with the LMC and
SMC distance moduli measurements given by Van den Bergh~(\cite{Vandenbergh00}),
who derived the mean values of distance moduli $\overline{(m - M)}_{0} = 18.50\pm0.05$ for LMC
and $\overline{(m - M)}_{0} = 18.85\pm0.1$ for SMC, corresponding to the heliocentric distances of
$\mathrm{(50.1\pm1.2)\,kpc}$ and $\mathrm{(58.9\pm2.6)\,kpc}$, respectively.
Only 2 of the 6 components of the LMC and SMC position vectors enter the GA search as
free parameters because the rest of them are determined by the projected position of the Clouds on the plane of sky, that
is $l_{LMC} = 280^{\circ}\,27'$, $b_\mathrm{LMC} = -32^{\circ}\,53'$ and
$l_\mathrm{SMC} = 302^{\circ}\,47'$, $b_\mathrm{SMC} = -44^{\circ}\,18'$.

Previous studies by Murai\,\&\,Fujimoto~(\cite{Murai80}) and Gardiner
et al.~(\cite{Gardiner94}) found that the correct choice of the
spatial velocities is crucial for reproducing the observed H\,I
structures. Kroupa\,\&\,Bastian~(\cite{Kroupa97}) derived proper motions of LMC and SMC stars from an analysis of
HIPPARCOS such that
$\mathrm{\vec{\upsilon}_{LMC} = (+41\pm44, -200\pm31,
  +169\pm37)\,km\,s^{-1}}$ and $\mathrm{\vec{\upsilon}_{SMC} =
  (+60\pm172, -174\pm172, +173\pm128)\,km\,s^{-1}}$.
The large uncertainties in actual values of the LMC and SMC velocity vectors originate in the fact that
the measured transverse velocities suffer
from large rms errors, which is connected to the large distance of LMC and
SMC in combination with the limited precision of proper motions in the
HIPPARCOS catalog.
To derive the actual initial conditions at the starting time of the simulation
from the current positions and velocities of the Clouds,
we employed the backward integration of equations of motion.

Current total masses $m_\mathrm{LMC}$ and $m_\mathrm{SMC}$ follow estimates by Van den Bergh~(\cite{Vandenbergh00}).
In general, masses of the Clouds are functions of time and evolve due to the LMC--SMC exchange of matter, and as a consequence of
the interaction between the Clouds and MW. Our test--particle model does not allow for a reasonable treatment of a time--dependent
mass--loss. Therefore, masses of the Clouds are considered constant in time, and their initial values at the starting epoch
of our simulations are approximated by the current LMC and SMC masses. Regarding the large range of the LMC and SMC
mass estimates available (for details see Van den Bergh~(\cite{Vandenbergh00})), $m_\mathrm{LMC}$ and $m_\mathrm{SMC}$ are also
treated as free input parameters of our model that become subjects to the GA optimization. Their ranges that we investigated can be found
in Table~\ref{tab_parameters}.

Scale radii of the LMC and SMC Plummer potentials $\epsilon_\mathrm{LMC}$ and $\epsilon_\mathrm{SMC}$
are input parameters of the model describing the radial mass distribution in the Clouds.
The study by Gardiner et al.~(\cite{Gardiner94}) used the values $\epsilon_\mathrm{LMC} = 3$\,kpc and
$\epsilon_\mathrm{SMC} = 2$\,kpc. To investigate the influence of this parameter on the evolution of the
Magellanic System, the Plummer radii were involved
in the GA search and their values were varied within the ranges of the width of 1\,kpc, including the estimates by Gardiner
et al.~(\cite{Gardiner94}) (see Table~\ref{tab_parameters}).

Gardiner et al.~(\cite{Gardiner94}) analyzed the H\,I surface contour
map of the Clouds to estimate the initial LMC and SMC disk radii
entering their model of the Magellanic System as initial conditions.
Regarding the absence of a clearly defined disk of the SMC, and
possible significant mass redistribution in the Clouds during their
evolution, the results require careful treatment and further
verification. We varied the current estimates of disk radii
$r^\mathrm{disk}_\mathrm{LMC}$ and $r^\mathrm{disk}_\mathrm{SMC}$ as free
parameters within the ranges introduced in Table~\ref{tab_parameters},
containing the values derived by Gardiner et al.~(\cite{Gardiner94}), and used them as
initial values at the starting point of our calculations ($T=-4$\,Gyr).

The orientation of the disks is described 
by two angles $\Theta$ and $\Phi$ defined by 
Gardiner\,\&\,Noguchi~(\cite{Gardiner96}).
Several observational determinations of the LMC disk plane orientation 
were collected by Lin et al.~(\cite{Lin95}).
Its sense of rotation is assumed to be clockwise (Lin et al.~\cite{Lin95}, 
Kroupa\,\&\,Bastian~\cite{Kroupa97}). 
The position angle of the bar structure in the SMC was used by 
Gardiner\,\&\,Noguchi~(\cite{Gardiner96}) to investigate the 
current disk orientation. Their results allow for wide ranges of the LMC and SMC disk
orientation angles (see Table~\ref{tab_parameters}) and so we investigated $\Theta$ and $\Phi$
by the GA search method, too. Similarly to Gardiner\,\&\,Noguchi~(\cite{Gardiner96}), we use the current LMC and
SMC disk orientation angles (Table~\ref{tab_parameters}) to approximate their initial values at $T=-4$\,Gyr.

\section{Results}
We try to reproduce as closely as possible the column density 
and velocity 
distribution of H\,I in the Magellanic Stream and in the 
Leading Arm. The influence of actual distances to the LMC and SMC and of 
their present space velocity vectors is considered together with their masses 
and the past sizes and space orientations of the orginal disks.
Here, we give the results of the search in the parameter space with the GA
using the 3--component $FF$ defined by Eq.~\ref{fitnessA}.
In principle, the GA is able to find the global extreme of the FF if enough time
is allowed for the evolution of the explored system (see Holland~\cite{Holland75}).
However, it may be very time--consuming to identify the single best fit due to a slow convergence of 
the FF. Therefore, to keep the computational cost reasonable, 
the maximum number of 120 GA generations to go through was defined.

To explore the $FF$ of our system, we collected 123 GA fits of the Magellanic System resulting from repeated runs of
our GA optimizer. Typically, identification of a single GA fit requires $\mathrm{\approx 10^4}$ runs of the numerical model.
Thus, due to the application of GA we were able to search the extended parameter space of the interaction
and discover the most successful models of the System over the entire parameter space by testing
$\mathrm{\approx 10^6}$ parameter combinations. In the case of our \mbox{20--dimensional} parameter space, simple exploration of every possible
combination of parameters even on a sparse grid of, e.g., 10 nodes per dimension, means $\mathrm{10^{20}}$ runs of the model.
Such a comparison clearly shows the necessity of using optimization techniques and demonstrates computational efficiency of
GA.
\begin{figure}[h]
\includegraphics[bb=45 65 430 675, angle=90, width = 8.7cm, clip]{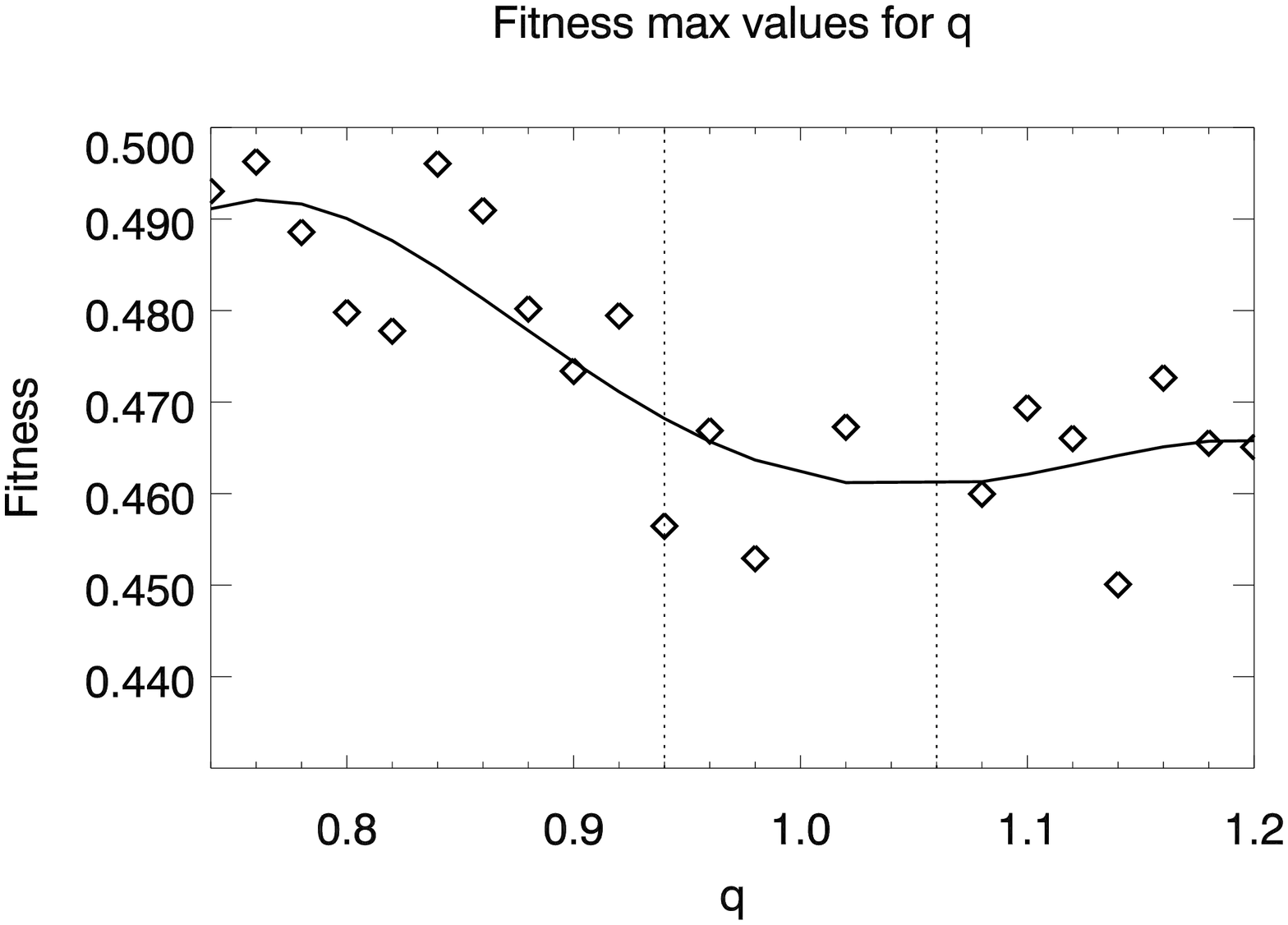}
\caption{Maximum values of fitness as a function of the MW dark matter halo flattening $q$.
The plot depicts the fitness of the best GA fit of the Magellanic System
that was found for each of the MW dark matter halo flattening values that entered the GA search.
The function is also approximated by its least--square polynomial fit.
The values of $q$ delimiting the model groups A, B, and C (see Table~\ref{tab_stat_q}) are emphasized by dotted lines.}
\label{pic_fitness-q_max}
\end{figure}

The fitness distribution for different values of the halo flattening parameter $q$ is shown in
Fig.~\ref{pic_fitness-q_max}. For every value of $q$, the model of the highest fitness is selected and its fitness value is
plotted. Figure~\ref{pic_fitness-q_max} indicates that better agreement between the models and observational data was generally achieved
for oblate halo configurations rather than for nearly spherical or prolate halos. 
The relative difference between the worst
($Fit_\mathrm{MIN} = 0.450; q=1.14$) and the best ($Fit_\mathrm{MAX} = 0.496; q=0.84$) model shown in Fig.~\ref{pic_fitness-q_max}
is $\Delta Fit = 0.09$. It reflects the fact that each of the GA fits contains a trailing tail (Magellanic Stream)
and a leading stream (Leading Arm), but that there are fine differences between the resulting distributions of matter. One may note
that the GA optimizer did not discover a model of fitness over 0.5 (the maximum reachable fitness value is 1.0 -- see Appendix~\ref{appendixB}).
It is caused either by insufficient volume of the studied parameter space of the interaction, or by simplification of physical
processes in our model (see Sec.~\ref{missing_stuff}), or by a combination of both reasons. That establishes an interesting
problem and should become a subject to future studies.

We want to discuss our results with 
respect to the shape of the MW halo.
\begin{table}[h]
\caption{Three major groups according to the halo flattening $q$}             
\label{tab_stat_q}      
\centering                          
\begin{tabular}{lccc}        
\hline                 
\hline                 
Group & A & B & C \\
\hline
 & $0.74 \leq q \leq 0.92$ & $0.94 \leq q \leq 1.06$ & $1.08 \leq q \leq 1.20$ \\    
\hline                                   
$N_i$ & 101 & 10 & 12\\    
\hline                                   
\end{tabular}
\end{table}
Thus, all the GA fits are divided into three
groups according to the halo flattening (see Table~\ref{tab_stat_q}) to show differences or
common features of models for oblate, nearly spherical, and prolate halo configurations.
The actual borders between the groups A, B, and C were selected by definition and, so that the number of models
in each of the groups allows for statistical treatment of the LMC and SMC orbital features and particle redistribution
that will be introduced in Sect.~\ref{scenarios}.
\subsection{Evolution}\label{scenarios}
\begin{figure*}
\centering
\includegraphics[bb=0 40 470 690, angle=90, width = 5.7cm, clip]{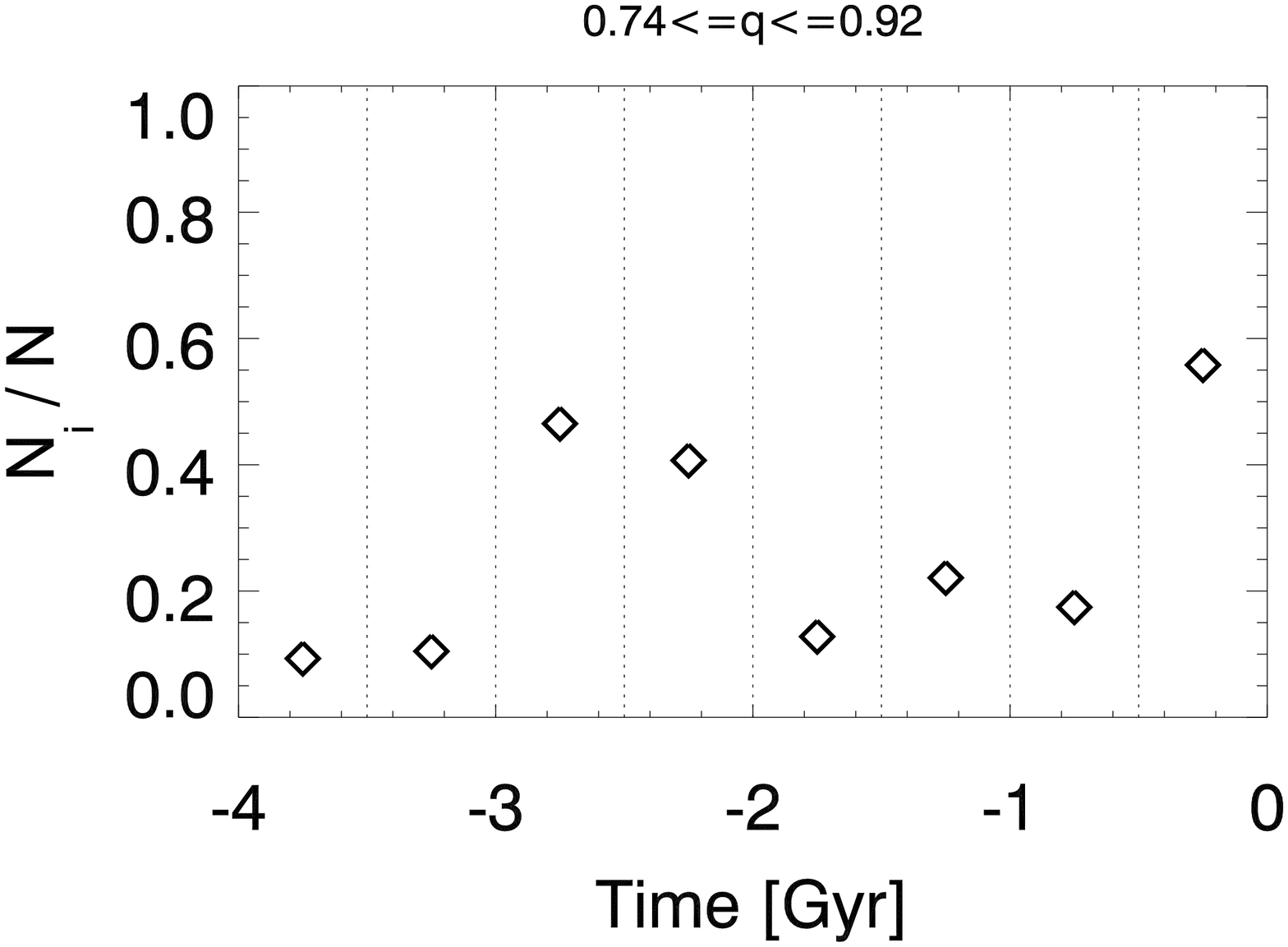}
\includegraphics[bb=0 40 470 690, angle=90, width = 5.7cm, clip]{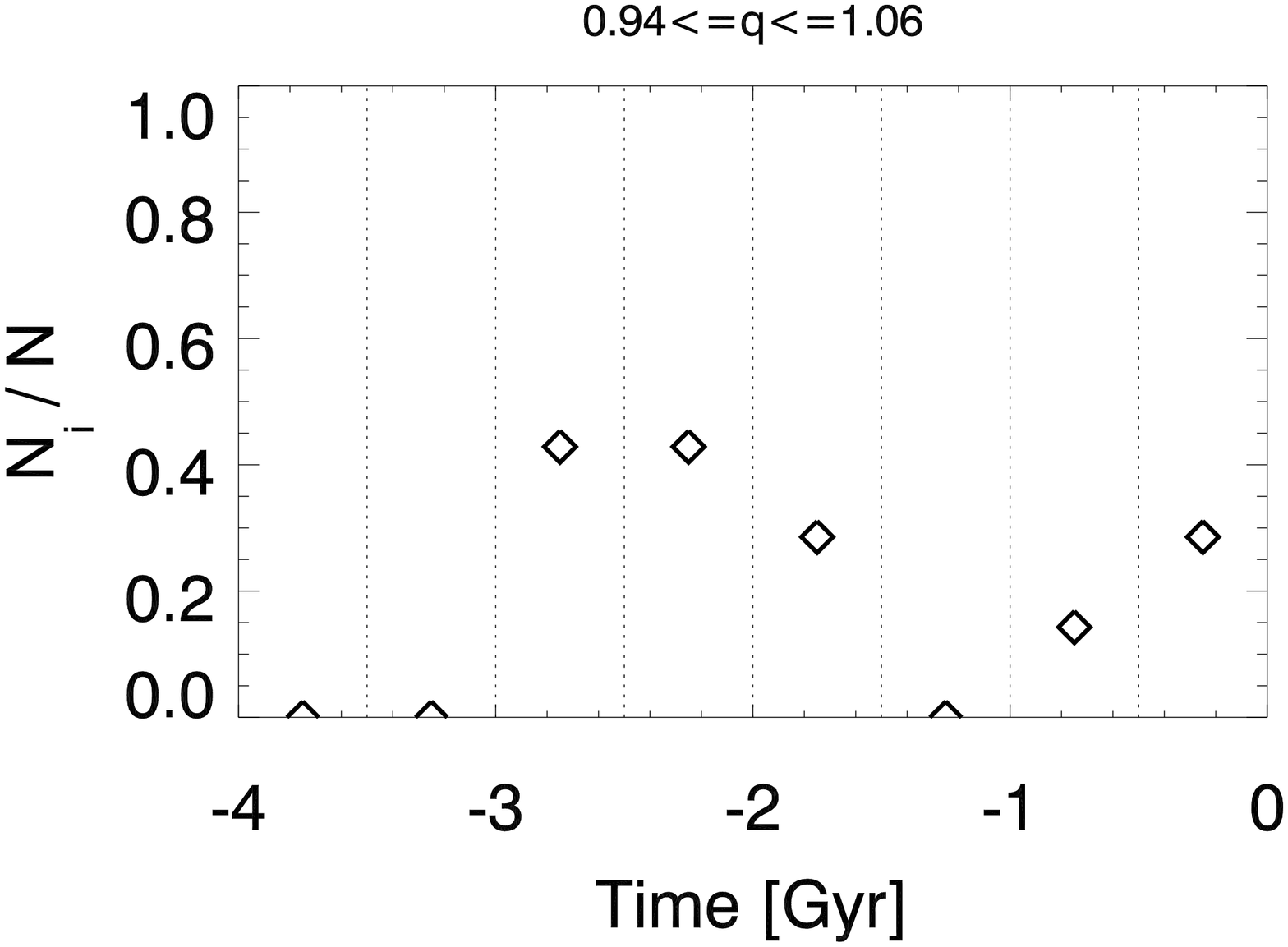}
\includegraphics[bb=0 40 470 690, angle=90, width = 5.7cm, clip]{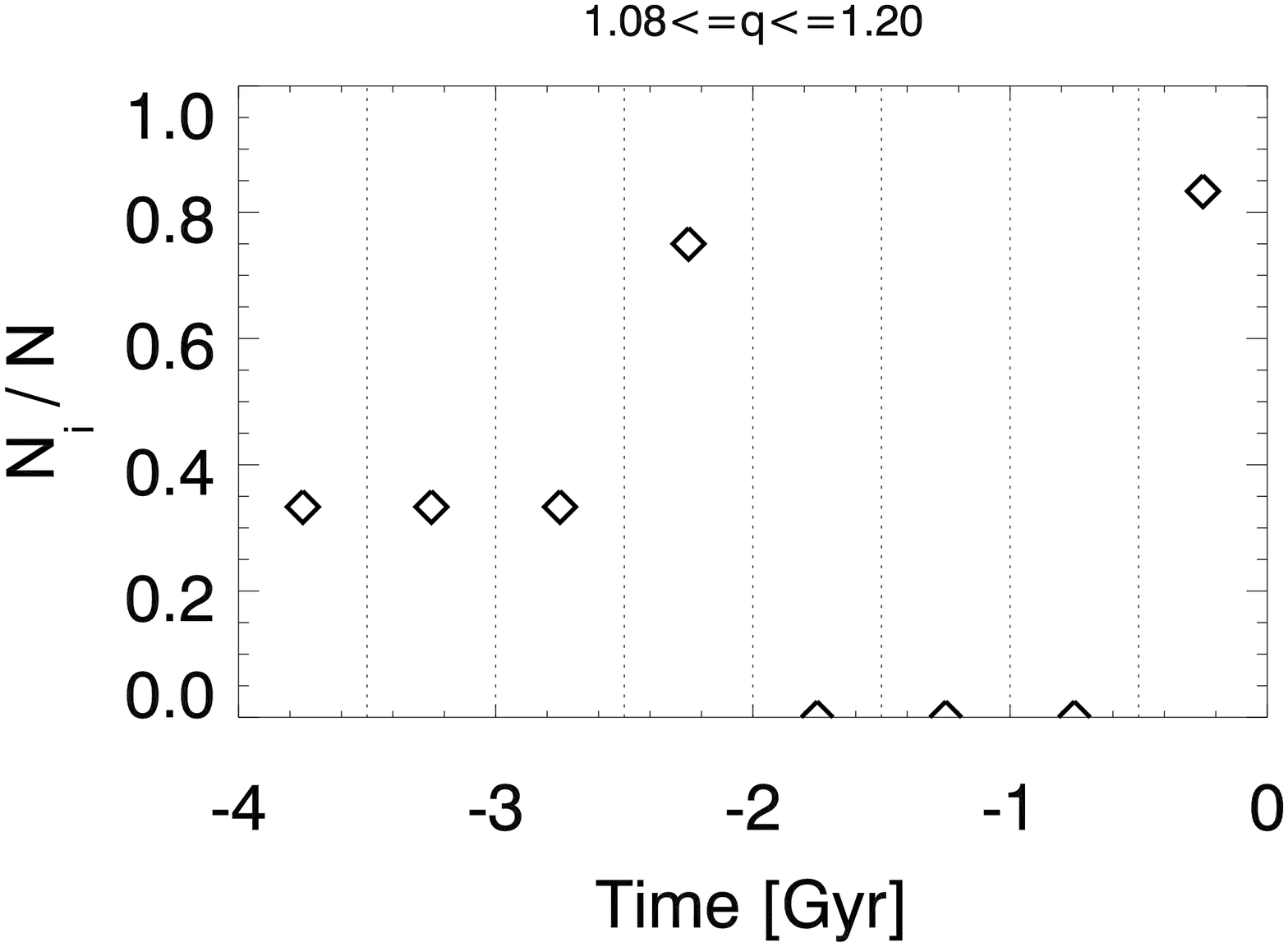}
\caption{Relative numbers of the Magellanic System GA models with
LMC--SMC relative distance minima at a given time interval for 
model groups A (left), B (middle), and C. 
The counts were made for 8 time intervals of 500\,Myr covering the entire Magellanic evolution period of 4\,Gyr 
investigated in our study.}
\label{pic_lsmin_stat-t}
\end{figure*}
\begin{figure*}
\centering
\includegraphics[bb=0 40 470 700, angle=90, width = 5.7cm, clip]{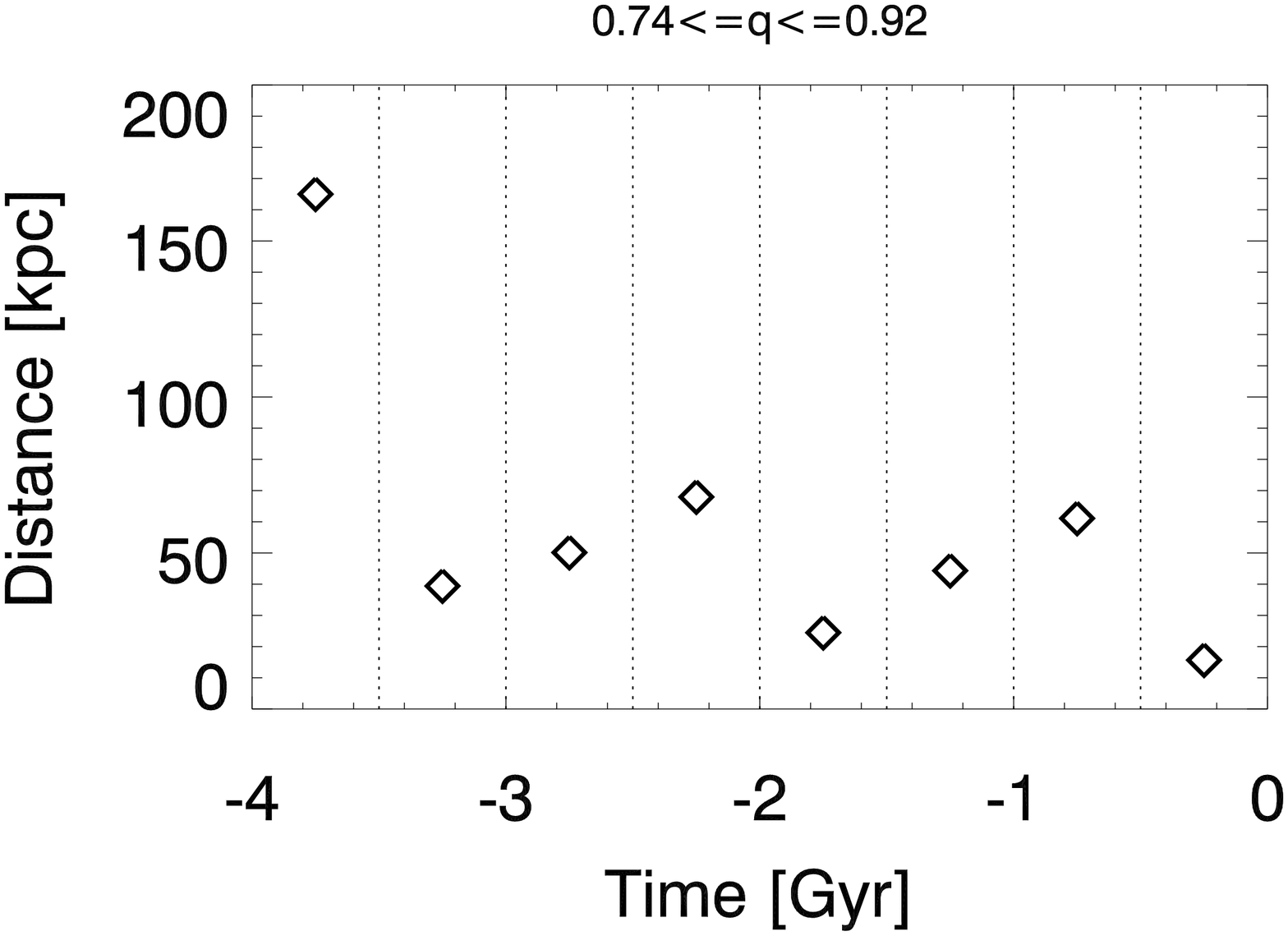}
\includegraphics[bb=0 40 470 700, angle=90, width = 5.7cm, clip]{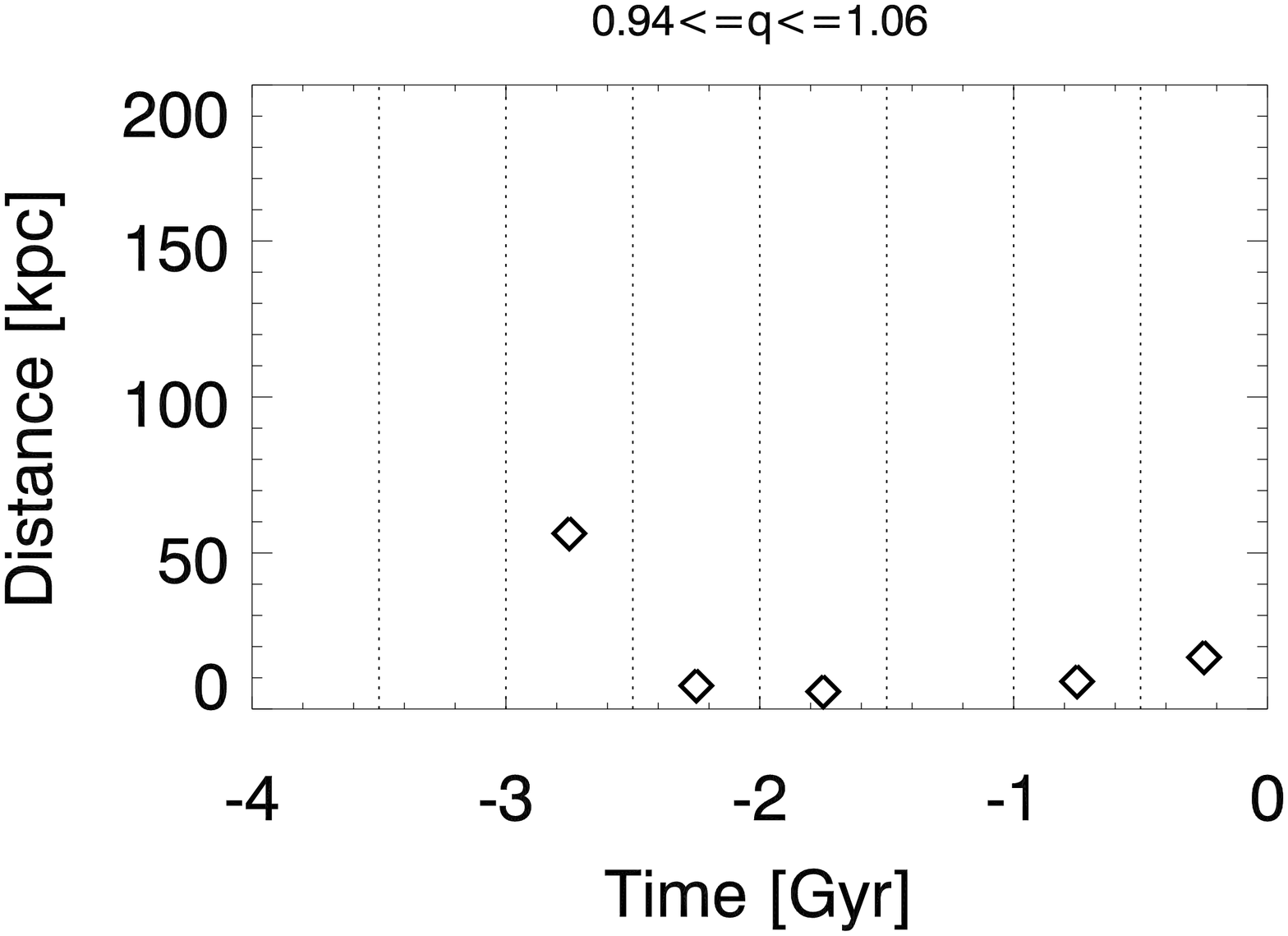}
\includegraphics[bb=0 40 470 700, angle=90, width = 5.7cm, clip]{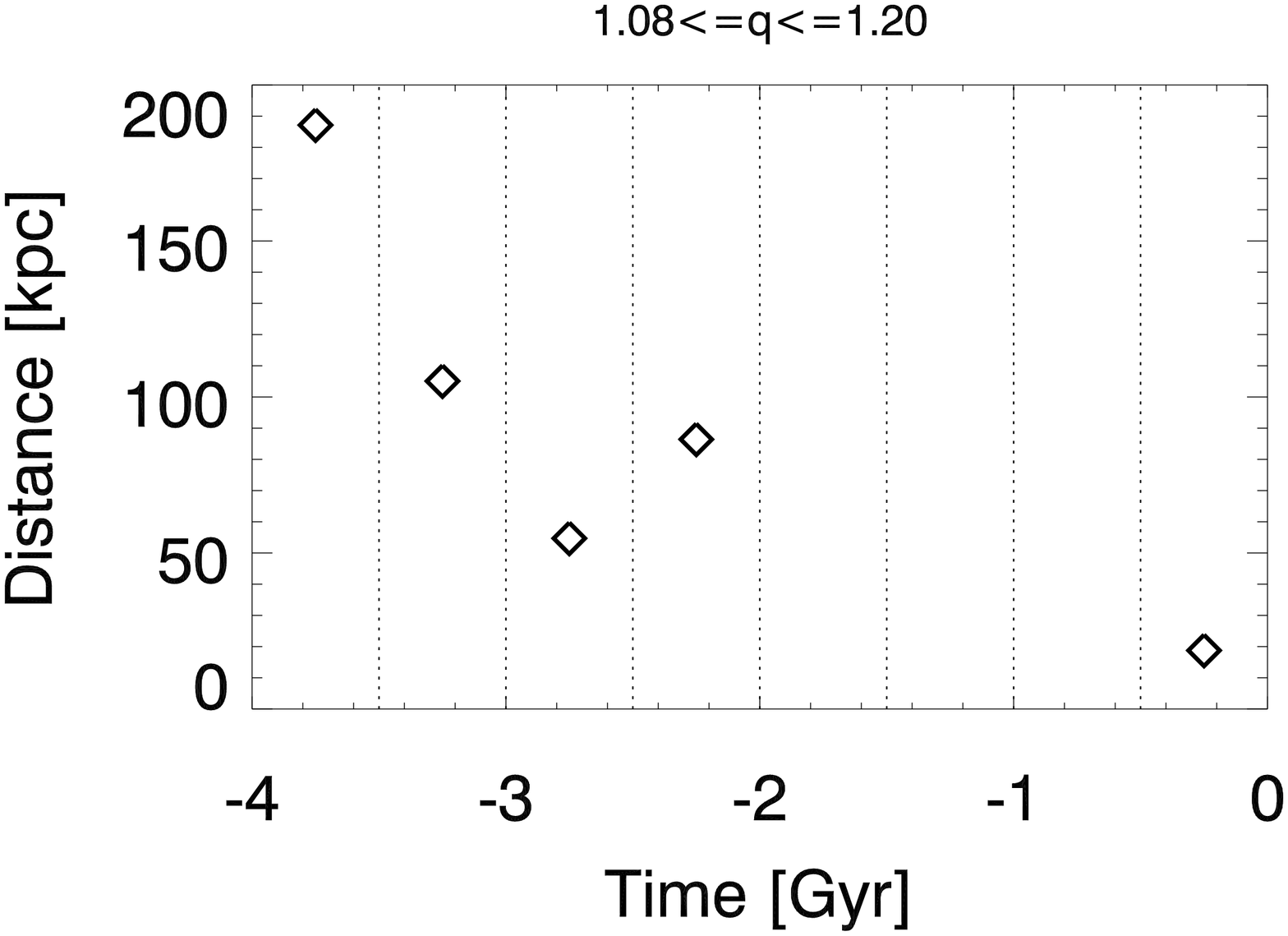}
\caption{Mean values of the LMC--SMC distance minima for model groups A (left), B (middle),
and C (Table~\ref{tab_stat_q}) and for 8 time intervals of 500\,Myr.}
\label{pic_lsmin_r-t}
\end{figure*}
\begin{figure*}
\centering
\includegraphics[bb=0 40 470 700, angle=90, width = 5.7cm, clip]{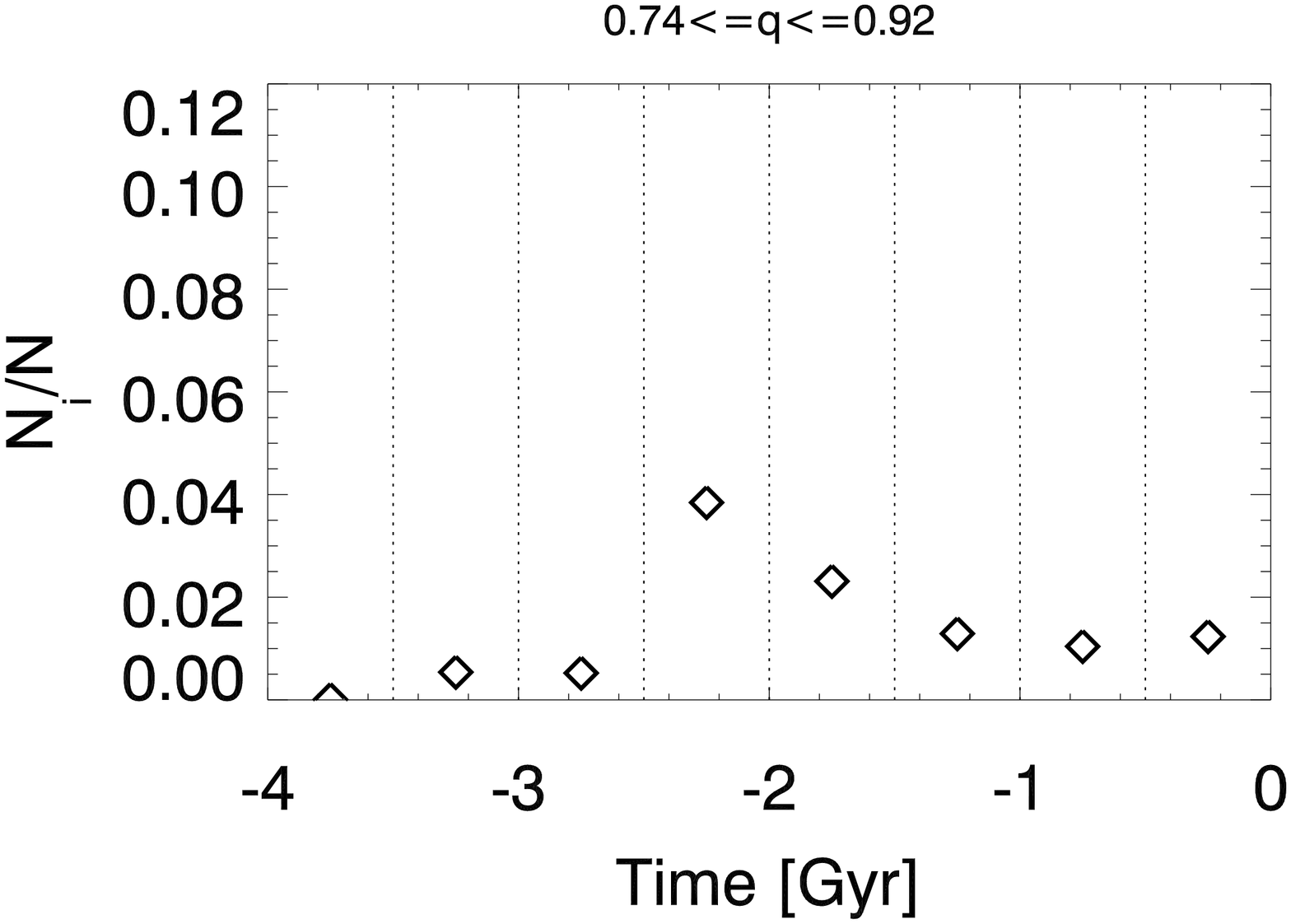}
\includegraphics[bb=0 40 470 700, angle=90, width = 5.7cm, clip]{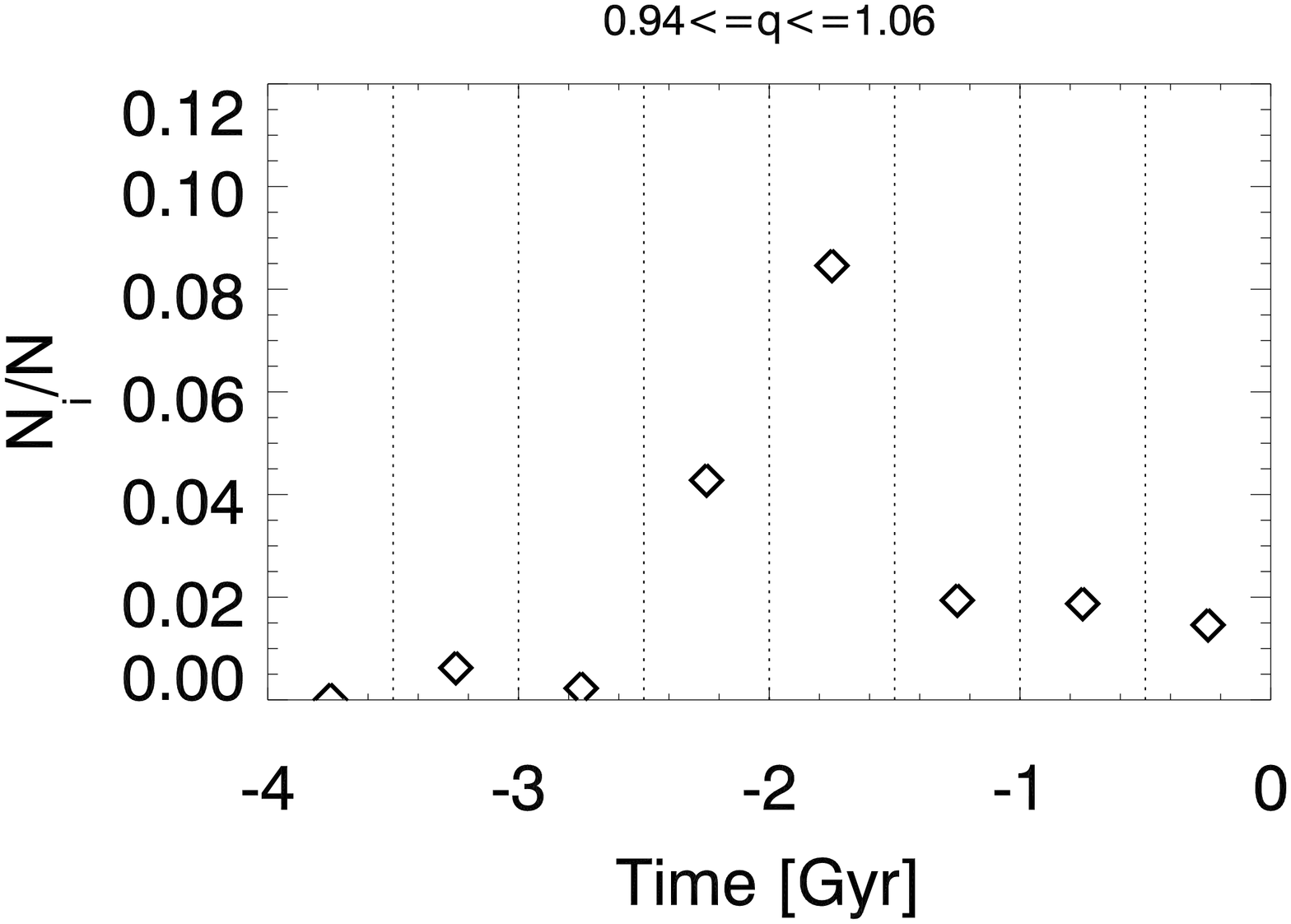}
\includegraphics[bb=0 40 470 700, angle=90, width = 5.7cm, clip]{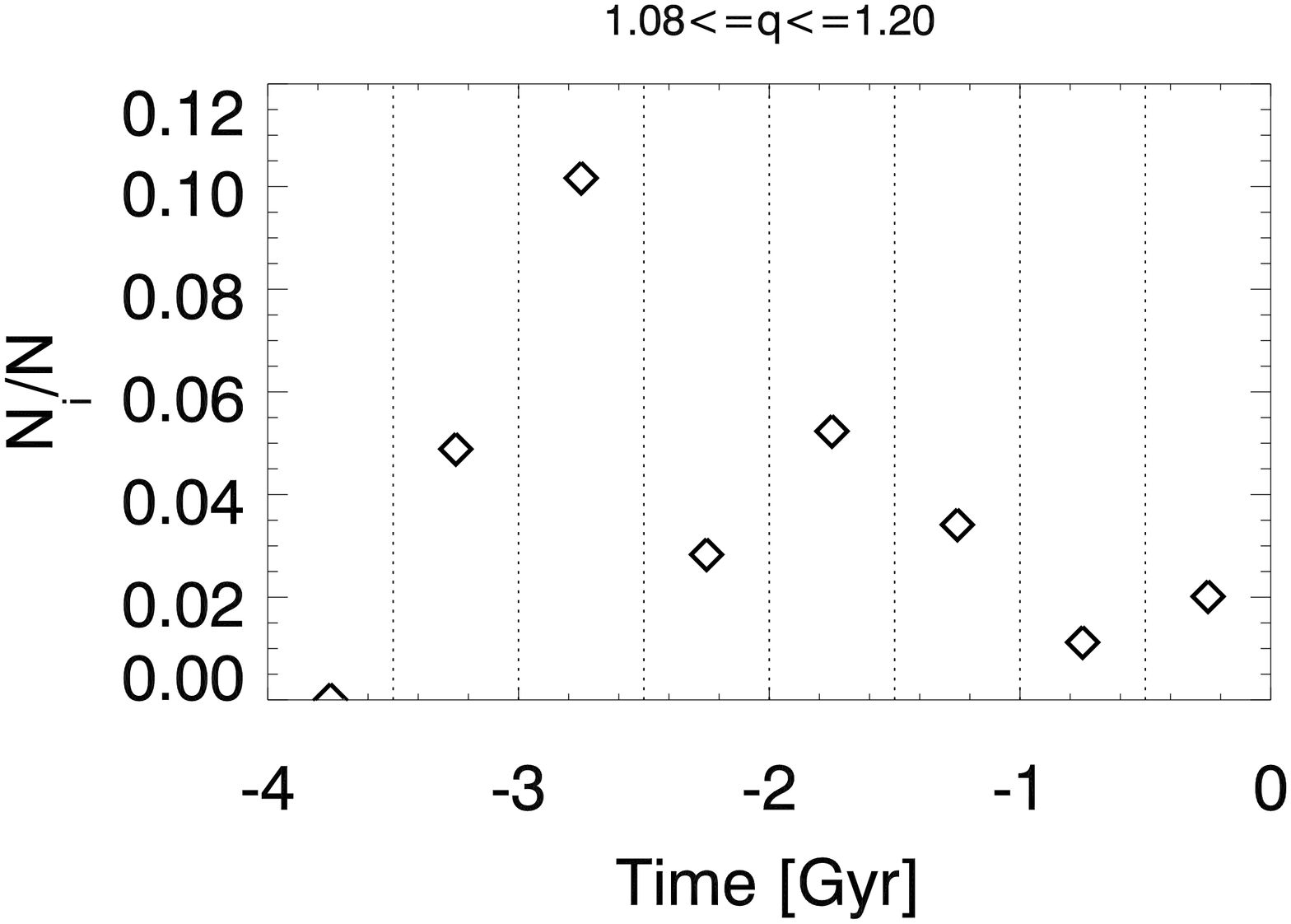}
\caption{Relative number of the LMC and SMC test--particles strongly disturbed due to the 
Magellanic System interaction. Counts within 8 time intervals of 500\,Myr and for
model groups A (left), B (middle), and C are plotted.}
\label{pic_lostparticles_relnum-t}
\end{figure*}

Close encounters of interacting galaxies often lead to substantial 
disruption of their particle 
disks, forming subsequently tidal arms and tails
(Toomre\,\&\,Toomre~\cite{Toomre72}). 
Regarding that, the time dependence of the relative distance of the interacting pair 
is an interesting source of information about the system.

First of all, we examine the time distribution of the minima of the LMC--SMC relative distance. 
For each of the model groups mentioned in Table~\ref{tab_stat_q},
we calculate the relative number of GA solutions having a minimum of the LMC--SMC relative distance
within a given interval of 500\,Myr. Figure~\ref{pic_lsmin_stat-t} shows such a distribution of fits for 
the total time interval of our simulations, which is 4.0\,Gyr.
The local maxima of the time distribution of the LMC--SMC distance minima are within the
intervals $\mathrm{\langle-3.0, -2.0\rangle\,Gyr}$ and $\mathrm{\langle-0.5, 0.0\rangle\,Gyr}$.
For prolate halos ($q \geq 1.08$) there is no LMC--SMC distance 
minimum between -2.0\,Gyr and -0.5\,Gyr.

Subsequently, the mean values of the LMC--SMC distance minima are calculated for each of the time 
intervals defined above. Comparison of the results for oblate, nearly spherical, and prolate DM halo 
configuration is available in Fig.~\ref{pic_lsmin_r-t}. It was found that close ($\Delta r \approx 10$\,kpc) LMC--SMC
encounters do not occur in models with either
oblate or prolate halos. If the MW DM halo shape is nearly spherical, disruption of the LMC and SMC initial particle
distribution leading to creation of the observed H\,I structures typically occurred due to strong LMC--SMC interaction.

Another point of interest is the time dependence of the LMC and SMC test--particle redistribution 
during the evolutionary process. Figure~\ref{pic_lostparticles_relnum-t} offers the relative number of 
test--particles strongly disturbed, i.e., particles that reached the minimal distance of twice the original radii of their circular
orbits around the LMC and SMC centers, respectively,
and by the LMC--SMC--MW interaction in the defined time--intervals.
Comparison between the plots in Figs.~\ref{pic_lsmin_r-t} and~\ref{pic_lostparticles_relnum-t}
shows that encounters of the Clouds are followed by delayed 
raise of the number of particles shifted to different orbits, typically. Another such events 
are induced by the interaction of the Clouds and MW.
Disruption of the LMC and SMC disks triggers formation of the extended structures of the Magellanic System.
Particles are assigned new orbits in the superimposed gravitational potential of the LMC, SMC, and MW, and spread
along the orbital paths of the Clouds.
Our study shows that the formation of the Magellanic Stream and other observed H\,I features did not begin earlier
than 2.5\,Gyr ago for model groups A and B (see Fig.~\ref{pic_lostparticles_relnum-t}). 
Prolate halos (group C)
allow for a mass redistribution in the system that started at $T < -3.5$\,Gyr.

\subsection{Representative models}
Here, we describe the models of highest fitness selected from each
of the groups A, B, and C. All of them are typical representatives of their model groups and we discuss
\begin{table}[h]
\centering                          
\caption{Parameters of the best models in separate $q$ categories.}             
\label{tab_sel_q}      
\begin{tabular}{lrrr}        
\hline                 
\hline                 
Model & A & B & C \\    
\hline                        
$q$ & \textbf{0.84} & \textbf{1.02} & \textbf{1.16} \\      
\hline                        
$Fit$ & 0.496 & 0.467 & 0.473 \\      
$\mathrm{\vec{r}_{LMC}[kpc]}$ &
$\left(\begin{array}{r} -1.26 \\ -40.50 \\ -26.87\end{array}\right)$ &
$\left(\begin{array}{r} -0.90 \\ -40.31 \\ -26.88\end{array}\right)$ &
$\left(\begin{array}{r} -0.63 \\ -40.03 \\ -26.92\end{array}\right)$ \\
$\mathrm{\vec{r}_{SMC}[kpc]}$ &
$\left(\begin{array}{r} 13.16 \\ -34.26 \\ -39.77\end{array}\right)$ &
$\left(\begin{array}{r} 13.32 \\ -34.33 \\ -40.22\end{array}\right)$ &
$\left(\begin{array}{r} 13.92 \\ -34.04 \\ -39.86\end{array}\right)$ \\
\hline                        
$\mathrm{\vec{\upsilon}_{LMC}[km\,s^{-1}]}$ &
$\left(\begin{array}{r} 44.0 \\ -169.8 \\ 146.7\end{array}\right)$ &
$\left(\begin{array}{r} 18.5 \\ -169.3 \\ 171.3\end{array}\right)$ &
$\left(\begin{array}{r}  5.8 \\ -169.2 \\ 205.8\end{array}\right)$ \\
$\mathrm{\vec{\upsilon}_{SMC}[km\,s^{-1}]}$ &
$\left(\begin{array}{r} -37.2 \\ -60.2 \\ 204.3\end{array}\right)$ &
$\left(\begin{array}{r} -10.1 \\ -94.2 \\ 270.0\end{array}\right)$ &
$\left(\begin{array}{r} -47.5 \\ -13.2 \\ 162.6\end{array}\right)$ \\
\hline                        
$m_\mathrm{LMC}[10^9\mathrm{M_\odot}]$ & 24.46 & 19.86 & 19.01 \\
$m_\mathrm{SMC}[10^9\mathrm{M_\odot}]$ & 2.06 & 1.82 & 1.83 \\
\hline                        
$r^\mathrm{disk}_\mathrm{LMC}\mathrm{[kpc]}$ & 9.62 & 11.46 & 9.06 \\
$r^\mathrm{disk}_\mathrm{SMC}\mathrm{[kpc]}$ & 6.54 & 6.06 & 7.90 \\
\hline                        
$\Theta^\mathrm{disk}_\mathrm{LMC}$ & $89^\circ$ & $98^\circ$ & $102^\circ$ \\
$\Phi^\mathrm{disk}_\mathrm{LMC}$ & $274^\circ$ & $277^\circ$ & $281^\circ$ \\
$\Theta^\mathrm{disk}_\mathrm{SMC}$ & $36^\circ$ & $49^\circ$ & $36^\circ$ \\
$\Phi^\mathrm{disk}_\mathrm{SMC}$ & $229^\circ$ & $231^\circ$ & $224^\circ$ \\
\hline
\end{tabular}
\end{table}
their features with respect to the H\,I observational data.
Table~\ref{tab_sel_q} summarizes the parameter values of the models.
\subsubsection{Group A}
\begin{figure}[h]
\centering
\includegraphics[bb=15 30 550 660, angle=90, width = 8.5cm, clip]{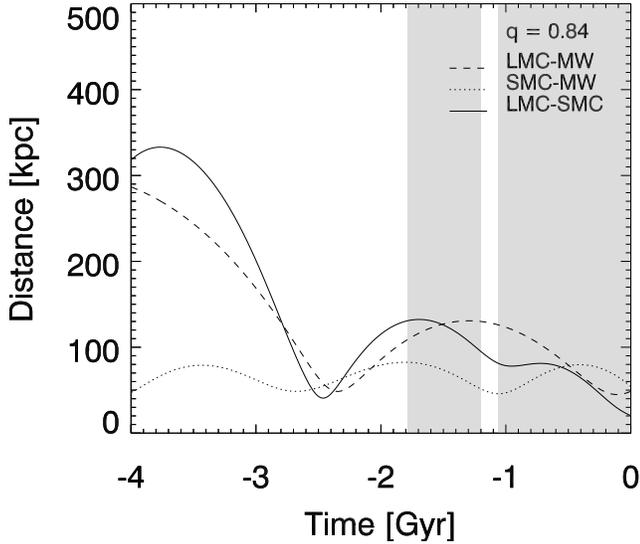}
\caption{Orbital evolution of the Clouds for the best GA fit from the model group A. The plot corresponds to an oblate halo of
flattening $q=0.84$. Time dependence of the LMC (dashed line) and
SMC (dotted line) galactocentric distances, and the LMC--SMC relative distance are plotted above. Plot areas with grey filling
mark the time intervals when the Clouds were gravitationally bound to each other.}
\label{pic_reldist_a}
\end{figure}

The best model of the Magellanic System with 
an oblate MW halo is introduced in this section (model A).
Figure~\ref{pic_reldist_a} depicts the time variation of the LMC and SMC 
galactocentric distances together with the LMC--SMC 
separation for the last 4\,Gyr. 
The Clouds move on very different orbits.
The apogalactic distance of LMC decreases systematically during 
the evolutionary period, which clearly reflects the effect of dynamical friction.
There is a gap between the periods of subsequent perigalactic approaches of the Clouds. While the last two perigalactica of LMC
are separated by $\mathrm{\approx 2.3\,Gyr}$, it is not over $\mathrm{\approx 1.5\,Gyr}$ in the case of the SMC.
Filled parts of the plot in Fig.~\ref{pic_reldist_a} indicate that the Magellanic Clouds have reached the state of
a gravitationally bound system during the last 4\,Gyr.
We define gravitational binding by the sum of the relative kinetic and gravitational potential energy of LMC and SMC.
The Clouds are bound when the total energy is negative.
Specifically, LMC and SMC have been forming a bound couple since $T = -1.06$\,Gyr.
Nevertheless, the total lifetime of a bound LMC--SMC pair did not exceed $\mathrm{40\,\%}$ of the entire evolutionary period we studied.

\begin{figure}[h]
\centering
\includegraphics[bb=10 70 430 705, angle=90, width = 8.5cm, clip]{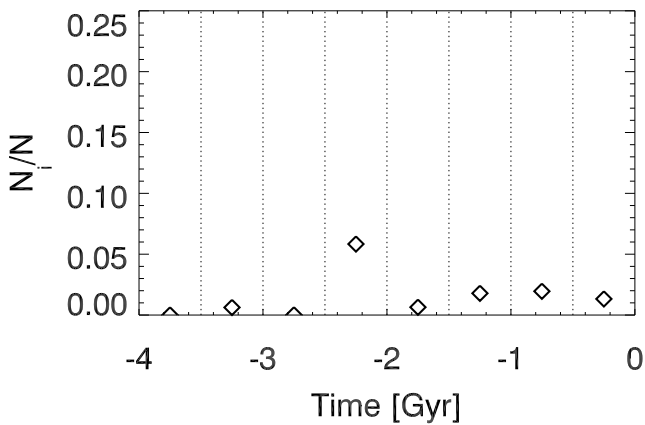}
\caption{Model A. Relative number of LMC/SMC test--particles strongly disturbed due to the interaction in the Magellanic System.
Counts within 8 time intervals of 500\,Myr are plotted.}
\label{pic_lostparticles_a}
\end{figure}

Comparison between Figs.~\ref{pic_reldist_a} and~\ref{pic_lostparticles_a} allows for conclusions about
major events that initialized the redistribution of the LMC and SMC particles. The most significant change of the initial
distribution of particles occurred as a result of the LMC--MW approach at $T \approx -2.4$\,Gyr, and the preceding
LMC--SMC encounter $T \approx -2.5$\,Gyr. Later on, the particle
redistribution continued due to tidal stripping by the MW.

\begin{figure}[h]
\centering
\includegraphics[bb=5 265 535 700, angle=90, width = 9.5cm, clip]{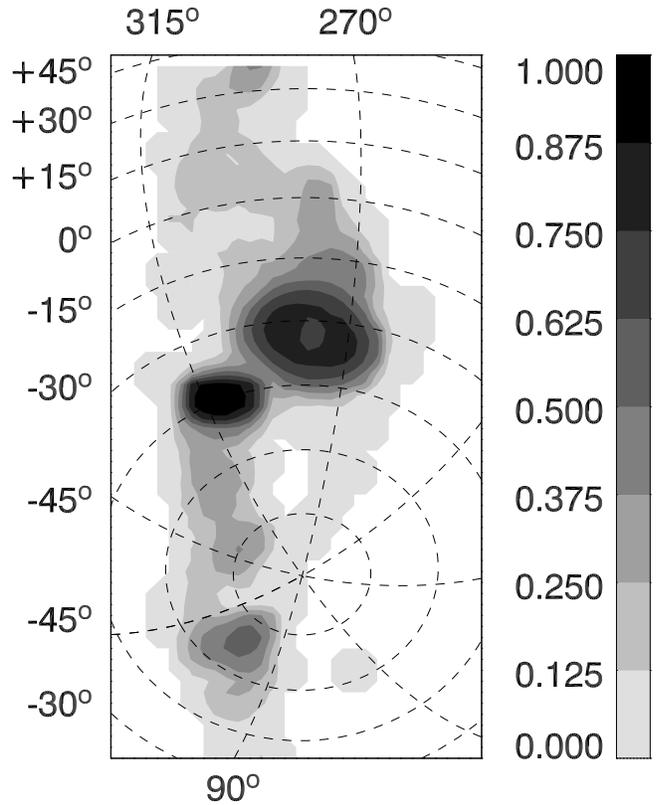}
\caption{Model A. Contour map of the modeled
H\,I integrated relative column density.
Data is projected on the sky plane. Galactic coordinates are used.}
\label{pic_mod_a1_dens}
\end{figure}

Figure~\ref{pic_mod_a1_dens} shows the modeled distribution of integrated H\,I column density
in the System. To enable comparison with the observed H\,I distribution,
we plotted a normalized H\,I column density map. The technique used to convert a test--particle distribution into a smooth
map of column densities is described in Appendix~\ref{appendixB}.
Mass distribution of H\,I extends beyond the far tip
of the observed Magellanic Stream (Fig.~\ref{pic_obs_dens}) in the model A. H\,I column density peaks can be found in Fig.~\ref{pic_obs_dens}
at the positions $l=300^{\circ}$, $b=-65^{\circ}$ and $l=45^{\circ}$, $b=-82^{\circ}$.
The model A places local density maxima of H\,I close to those observed ones
(i.e., relative angular distance is $\approx 10^{\circ}$)
to approximate positions $l=325^{\circ}$, $b=-70^{\circ}$
and $l=70^{\circ}$, $b=-70^{\circ}$, respectively.
Note also the low--density distribution of matter spread along the great circle of $l=270^{\circ}$ (Fig.~\ref{pic_mod_a1_dens}). The matter
emanates from LMC near the position of the \emph{Interface Region} identified by Br\"uns et al.~\cite{Bruens05} (Fig.~\ref{2d_HI_map}).
In general, the model overestimates the amount of matter in the Magellanic Stream.
The Leading Arm consists only of LMC particles in this scenario.
The modeled matter distribution covers a larger area of the plane of sky than what is observed. 
However, this is a common problem of previous test--particle models of the Magellanic System (see, e.g., Murai\,\&\,Fujimoto~\cite{Murai80},
Gardiner et al.~\cite{Gardiner94}) and is likely caused by simplifications in the treatment of the physical processes.
But also in general, successful reproduction of the Leading Arm
has been a difficult task for all the models introduced so far.

\begin{figure}[h]
\centering
\includegraphics[bb=30 100 472 670, angle=90, width = 8.5cm, clip]{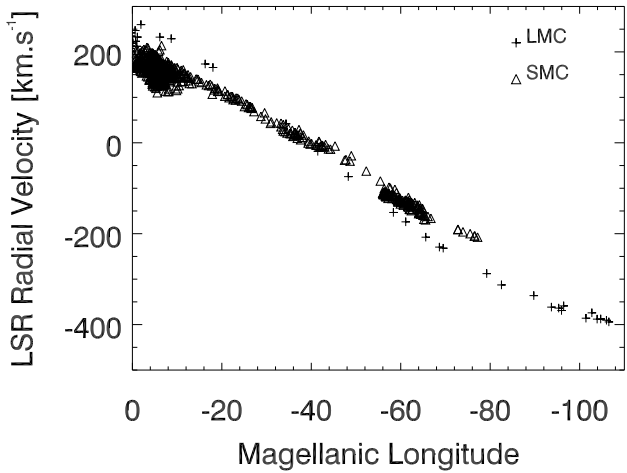}
\caption{Model A.
LSR radial velocity profile of the Magellanic Stream.}
\label{pic_mod_a1_radvel}
\end{figure}

The LSR radial velocity profile of the Magellanic Stream for the model A is shown in 
Fig.~\ref{pic_mod_a1_radvel}.
The model reproduces the LSR radial velocity of the Stream matter as an almost linear function of Magellanic Longitude with
the high--negative velocity tip reaching $\mathrm{-400\,km\,s^{-1}}$. Such features are in agreement
with observations (see Fig.~\ref{pic_obs_radvel}).
In contrast to Gardiner et 
al.~(\cite{Gardiner94}), the Magellanic Stream consists of both LMC and SMC particles. Figure~\ref{pic_mod_a1_radvel} denotes that the 
LMC and SMC Stream components cover different ranges of LSR radial velocities. The Stream component of the SMC origin does not extend
to LSR radial velocities below the limit of $\mathrm{-200\,km\,s^{-1}}$. The major fraction of the LMC particles resides
in the LSR radial velocity range from $\mathrm{-400\,km\,s^{-1}}$ to $\mathrm{-200\,km\,s^{-1}}$.

\subsubsection{Group B}
\begin{figure}[h]
\centering
\includegraphics[bb=15 30 550 660, angle=90, width = 8.5cm, clip]{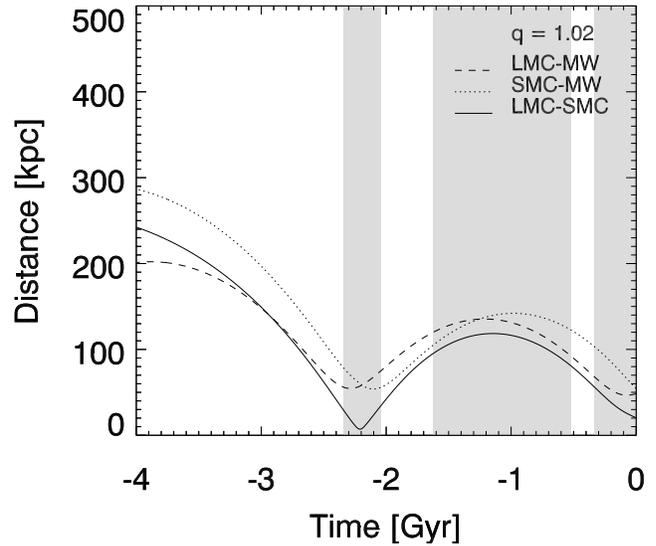}
\caption{Orbital evolution of the Clouds for the best GA fit from the model group B. The plot corresponds to a nearly spherical halo of
flattening $q=1.02$. Time dependence of the LMC (dashed line) and
SMC (dotted line) galactocentric distances, and the LMC--SMC relative distance are plotted above. Grey filled
areas show time intervals when the LMC and SMC formed a gravitationally bound couple.}
\label{pic_reldist_b}
\end{figure}

The best model of the group B (model B) corresponds to the MW DM halo flattening value $q=1.02$.
The initial condition set for the model B is listed in Table~\ref{tab_sel_q}.
The galactocentric distance of the Clouds and their spatial separation 
as functions of time are plotted in Fig.~\ref{pic_reldist_b}.
Continuous decrease of the LMC and SMC galactocentric distances due to the dynamical friction
is apparent for both LMC and SMC.
A very close encounter of the Clouds with the relative distance $\Delta r \approx 10$\,kpc
occurred at $T \approx -2.2$\,Gyr. At similar moments of $T \approx -2.3$\,Gyr (LMC) and
$T \approx -2.1$\,Gyr (SMC), the Clouds also reached perigalactica of their orbits. In general,
both Clouds have been moving on orbits showing similar time dependence of their galactocentric
distances, as indicated by  Fig.~\ref{pic_reldist_b}. Nevertheless, the position vectors of the Clouds
evolved in significantly different ways. As a consequence, the spatial separation of the Clouds varied within a wide
range of values from $\Delta r \approx 10$\,kpc to $\Delta r \approx 250$\,kpc.
The Clouds have formed a gravitationally bound couple three times within the last 4\,Gyr, and the total duration
of such periods was 1.7\,Gyr.
Currently, LMC and SMC are gravitationally bound in the model B. 

\begin{figure}[h]
\centering
\includegraphics[bb=10 70 430 705, angle=90, width = 8.5cm, clip]{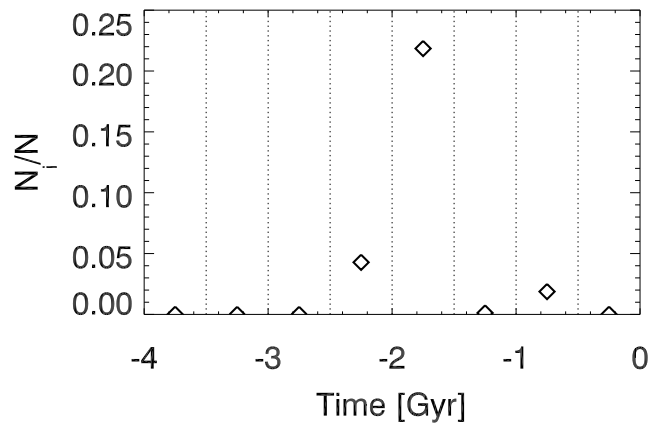}
\caption{Model B. Relative number of LMC/SMC test--particles strongly disturbed due to the interaction in the Magellanic System.
Counts within 8 time intervals of 500\,Myr are plotted.}
\label{pic_lostparticles_b}
\end{figure}

\begin{figure}[h]
\centering
\includegraphics[bb=5 265 535 700, angle=90, width = 9.5cm, clip]{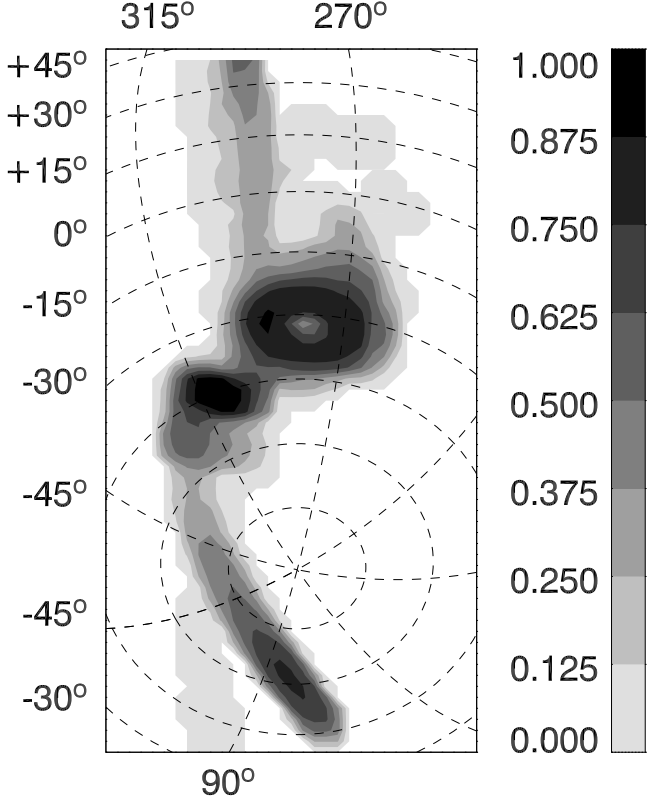}\\
\caption{Model B. Contour map of the modeled
H\,I integrated relative column density. Data is projected on the plane of sky. Galactic coordinates are used.
The dominant branch of the trailing stream is along the great circle denoted by
$l=135^{\circ}/315^{\circ}$.}
\label{pic_mod_b1_dens}
\end{figure}

The LMC--SMC encounter at $T \approx -2.2$\,Gyr caused a distortion of the original particle disks of the Clouds.
More than $\mathrm{25\,\%}$ of the total number of the LMC and SMC particles were moved to significantly different orbits (for a definition see
Sect.~\ref{scenarios}) within the interval of 1\,Gyr after the encounter (see Fig.~\ref{pic_lostparticles_b}).
The following evolution of the particle distribution formed extended structures depicted in Fig.~\ref{pic_mod_b1_dens}.
There are two spatially separated components present in the modeled tail.
The H\,I column density distribution map for the model B (see Fig.~\ref{pic_mod_b1_dens}) shows a densely populated trailing 
stream parallel to the great circle of $l=135^{\circ}/315^{\circ}$. 
It consists of the SMC particles torn off 
from the initial disk
$\mathrm{\approx 2\,Gyr}$ ago.
Its far end is projected to the plane of sky close to the tip of the Magellanic Stream.
However, the modeled increase of the column density of matter toward the far end of the tail is a substantial drawback of
scenario B.
The stream extends into the SMC leading arm located between $l=290^{\circ}, b=-15^{\circ}$ and
$l=290^{\circ}, b=45^{\circ}$. 
The second component of the particle tail is of LMC origin and is spread over the 
position of the observed low--density gas distribution centered at $l=80^{\circ}$, $b=-50^{\circ}$ (Figs.~\ref{2d_HI_map}
or~\ref{pic_obs_dens}).

The most significant structure at the leading side of the Magellanic System is the SMC stream introduced in the previous paragraph.
Comparison between Figs.~\ref{pic_obs_dens} and~\ref{pic_mod_b1_dens} indicates that neither the projected position nor the
\begin{figure}[h]
\centering
\includegraphics[bb=30 100 472 670, angle=90, width = 8.5cm, clip]{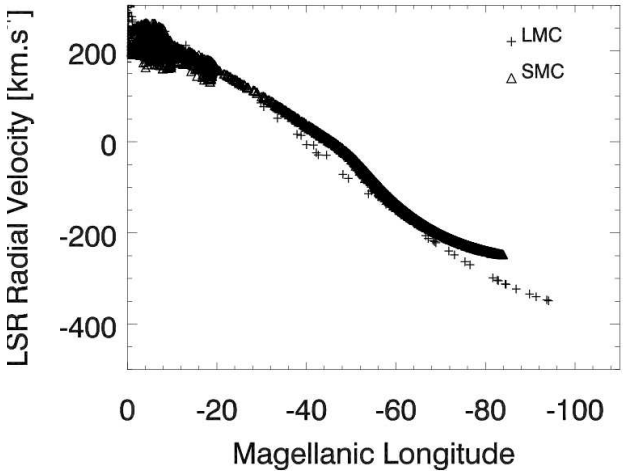}
\caption{Model B. LSR radial velocity profile of the Magellanic Stream.}
\label{pic_mod_b1_radvel}
\end{figure}
integrated H\,I density distribution of the stream is in agreement with the observed Leading Arm. There was also a structure emanating
from the leading edge of the LMC identified at approximate position $l=270^{\circ}, b=-15^{\circ}$ (see Fig.~\ref{pic_mod_b1_dens}).
Regarding the H\,I data by Br\"uns et al.~(\cite{Bruens05}),
such an H\,I distribution is not observed.
The LSR radial velocity profile of the trailing tail of the model B does not extend over
the limit of $\mathrm{\upsilon_{LSR} \approx -350\,km.s^{-1}}$
(Fig.~\ref{pic_mod_b1_radvel}). However, following the H\,I data, the far tip of the Stream should reach
the LSR radial velocity $\mathrm{\upsilon_{LSR} \approx -400\,km.s^{-1}}$ at the Magellanic Longitude
$\mathrm{\approx -100^{\circ}}$.

\subsubsection{Group C}
Our last model group C assumes the presence of prolate MW DM halos. The best GA fit 
of the System (model C) is 
introduced in Table~\ref{tab_sel_q} reviewing its initial condition set.
Concerning orbital motion of the Clouds, there is significant difference between the LMC and SMC periods of perigalactic
approaches during the last 4\,Gyr.
\begin{figure}[h]
\centering
\includegraphics[bb=15 30 550 660, angle=90, width = 8.5cm, clip]{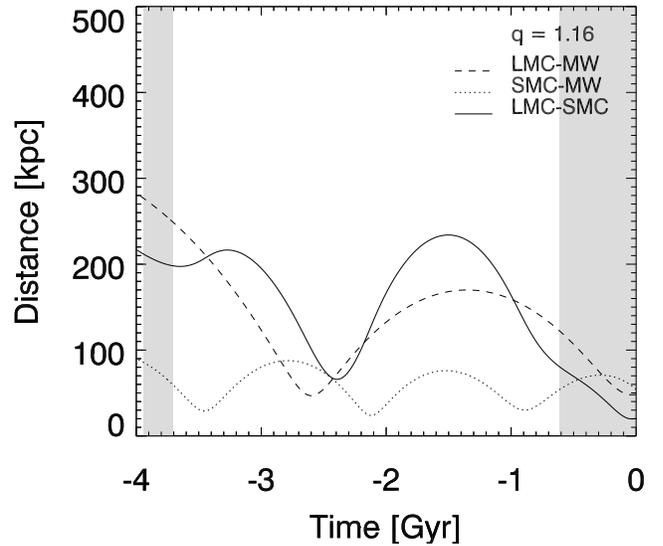}
\caption{Orbital evolution of the Clouds for the best GA fit from the model group C. The plot corresponds to a prolate halo of
flattening $q=1.16$. Time dependence of the LMC (dashed line) and
SMC (dotted line) galactocentric distances, and the LMC--SMC relative distance are plotted above. Periods when
the Clouds formed a gravitationally bound couple are marked by grey filling.}
\label{pic_reldist_c}
\end{figure}
While the last period of the LMC exceeds 2.5\,Gyr, the SMC orbital cycle is shorter 
than 1.5\,Gyr. The relative distance of the Clouds remains over 70\,kpc for $T<-0.4$\,Gyr. They became
a gravitationally bound couple 0.6\,Gyr ago and this binding has not been disrupted (Fig.~\ref{pic_reldist_c}).
\begin{figure}[h]
\centering
\includegraphics[bb=10 70 430 705, angle=90, width = 8.5cm, clip]{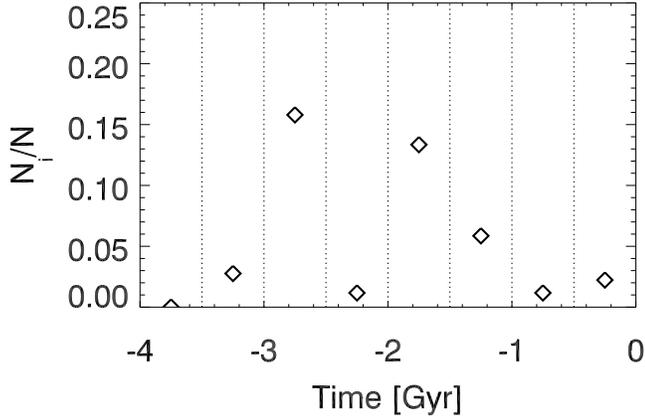}
\caption{Model C. Relative number of LMC/SMC test--particles strongly disturbed due to the interaction in the Magellanic System.
Counts within 8 time intervals of 500\,Myr are plotted.}
\label{pic_lostparticles_c}
\end{figure}

\begin{figure}[h]
\centering
\includegraphics[bb=5 265 535 700, angle=90, width = 9.5cm, clip]{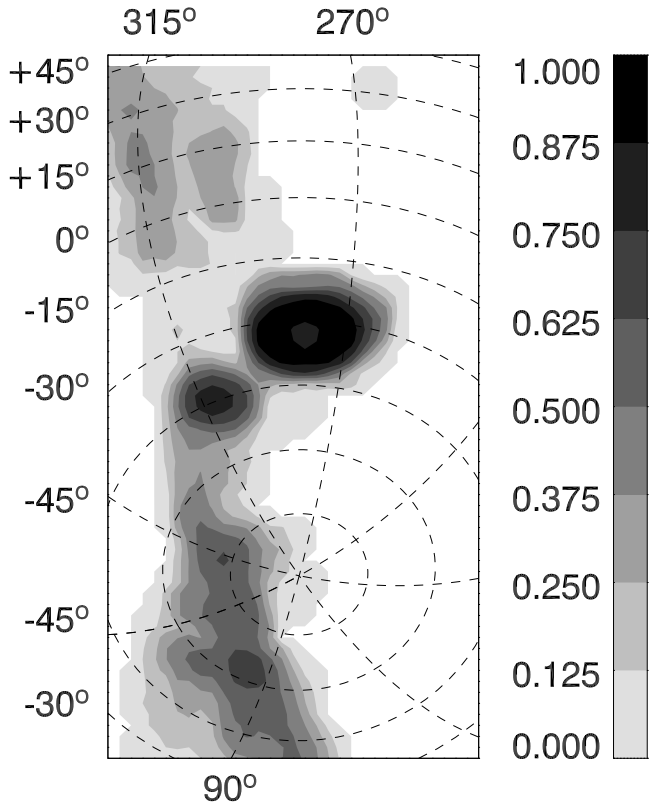}\\
\caption{Model C. Contour map of the modeled
H\,I integrated relative column density. Data is projected on the sky plane. Galactic coordinates are used.}
\label{pic_mod_c1_dens}
\end{figure}

Changes to the original LMC and SMC particle disks occurred especially due to the LMC--MW and SMC--MW encounters at 
$T<-2.0$\,Gyr. Comparison between Figs.~\ref{pic_reldist_c} and~\ref{pic_lostparticles_c} demonstrates
the significance of different encounter events for particle redistribution. Note that the rise of the number of disturbed
particles is delayed with respect to the corresponding disturbing event.
Subsequently, the evolution of particle structures continued under the influence of tidal stripping by the gravitational field of MW.
The current distribution of matter in the model C is plotted in the form of a 2--D map in Fig.~\ref{pic_mod_c1_dens}.
The projection of the modeled trailing stream indicates that it occupies larger area of the data cube
than the observed Magellanic Stream (compare Figs.~\ref{pic_obs_dens} and~\ref{pic_mod_c1_dens}).
According to Fig.~\ref{pic_obs_dens} the Magellanic Stream shows H\,I density peaks at $l=300^{\circ}$, $b=-65^{\circ}$ and $l=45^{\circ}$, $b=-82^{\circ}$.
Our model C expects two local maxima of H\,I integrated column density in the tail.
Their positions are shifted by $\mathrm{\approx 20^{\circ}}$ relatively to the peaks in Fig.~\ref{pic_obs_dens}.
Additional comparison between the model and observations reveals that the model C overestimates the integrated column densities
of H\,I in the Magellanic Stream.
The matter located at the leading side of the Magellanic System is of SMC origin only. Similarly to the case of
the trailing tail, the modeled amount of matter exceeds observational estimates for the Leading Arm. 

\begin{figure}[h]
\centering
\includegraphics[bb=30 100 472 670, angle=90, width = 8.5cm, clip]{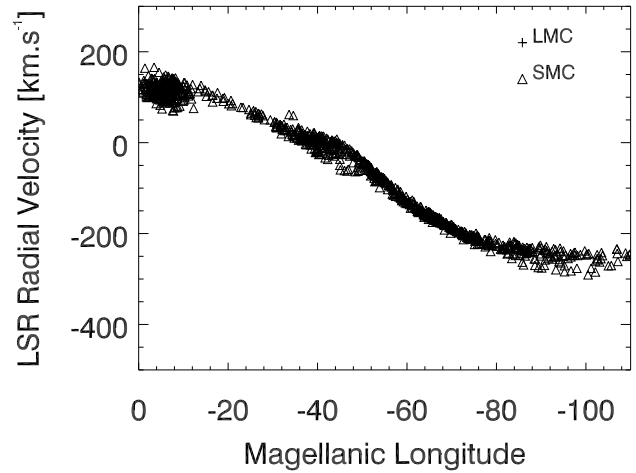}
\caption{Model C. LSR radial velocity profile of the Magellanic Stream.}
\label{pic_mod_c1_radvel}
\end{figure}

Figure~\ref{pic_mod_c1_radvel} offers the LSR radial velocity profile of the Magellanic Stream in our model C.
The measured minimum of the LSR radial 
velocity is $\mathrm{\approx -400\,km\,s^{-1}}$. The high negative LSR radial velocity at the far tip of the modeled Magellanic Stream 
does not exceed $\mathrm{\approx -300\,km\,s^{-1}}$, however.
The observed H\,I emission intensity decreases towards 
the high negative velocity tip, which was not well reproduced by the model C.


\section{Summary of the GA models}
\subsection{Orbits of the Magellanic Clouds}
Exploration of the orbital motion of the Clouds shows the similarity of the GA fits for oblate and prolate 
halos (models A and C). No close ($r<10$\,kpc) LMC--SMC 
encounters occurred for either of the models A and C.
It is also notable that in the models with aspherical halos, the SMC period of perigalactic approaches is significantly shorter than the period of the 
LMC and that the SMC remains closer than 100\,kpc to the MW center during the last 4\,Gyr. When the MW DM flattening 
$q \approx 1.0$, the LMC and SMC orbital cycle lengths were comparable for the model B. 
Independently of the MW halo shape, the LMC and SMC are currently forming a gravitationally bound couple in our models.
However, the Clouds cannot be considered bound to each other during the entire period of the last 4\,Gyr.
This is in contrast with Murai\,\&\,Fujimoto~(\cite{Murai80}), Gardiner et al.~(\cite{Gardiner94}), or 
Gardiner\,\&\,Noguchi~(\cite{Gardiner96}), who argue that the LMC and SMC moving in the spherical halo have formed
a gravitationally bound pair for at least several Gyr to allow for sufficient matter redistribution.
We show that the structural resolution adopted by the above--cited studies to make a comparison between models and observations
does not allow for such constraints of the orbital history of the Clouds. The GA search employed a 3--level detailed
evaluation of the modeled H\,I distribution with respect to high--resolution observational data (Br\"uns et al.~\cite{Bruens05}),
which introduced significant improvement of previous approaches to compare observations and models.
Nevertheless, there is still no clear indication that continuous gravitational binding of the Clouds covering the entire
evolutionary period is necessary
for successful reproduction of the observed data.

\subsection{Origin of the matter in the Stream}
In our best model both SMC and LMC particles were present in the trailing stream. This is a common feature of the scenarios
that were investigated. In general, the fraction of H\,I gas originating
at the SMC exceeds the fraction of LMC matter in the Stream.

Following the models A and B, the formation of the Magellanic Stream did not start
earlier than 2.5\,Gyr ago. In the case of the models C, the age of the Stream is $\mathrm{\approx 3.5\,Gyr}$.
The estimates by the models A and B are close to the epoch when a massive
star formation burst in LMC began (Van den Bergh~\cite{Vandenbergh00}). It is also true for the model C. However,
the disturbing event that triggered the formation of the Magellanic Stream in the model C occurred at a time
that does not substantially differ from the epoch $T=-4$\,Gyr when our simulation started. In such a case
we have to assume that the tidal interaction responsible for the creation of the Stream was very likely invalidating
the natural condition of sufficient isolation of the Clouds at the starting point of the simulation. Such a condition
is necessary to set up the initial particle disks around LMC and SMC when the particles reside on circular orbits.
Thus, the statements on the actual age of the Stream in the model C should be made with care. Regardless, the large
uncertainty of the estimated beginning of the LMC star burst (see Van den Bergh~\cite{Vandenbergh00}) requires
that consideration is given to the possibility that for case A, B, or C
the starting epochs of the Stream creation and of the
start formation burst in the LMC may overlap.

The previous paragraph indicates that it is very likely that matter forming the Stream comes from the Magellanic Clouds containing stars,
and we necessarily face the observational fact that
there is no stellar content in the Magellanic Stream. The models for aspherical halos
indicate that the matter coming to the Magellanic Stream from the LMC originates in the outer regions of its initial particle
disk, while no matter was torn off from the inner disk of radius $\mathrm{r_{disk} \approx 5\,kpc}$ that was the dominant
region of star formation in the LMC. It is due to the absence of close encounters in the Magellanic System.

In contrast to the models A and C, a dramatic encounter event
between the Clouds occurred in model B at $\mathrm{\approx
-2.2\,Gyr}$, when the internal structure of both disks was
altered and the matter from central areas of the LMC
disk was also transported to the Magellanic Stream.  In such a case we
expect a certain fraction of the matter of the Magellanic Stream to be
in the form of stars, which is, however, not supported by observations.

\subsection{Structure of the Stream}\label{stream_struct}
Mathewson et al.~(\cite{Mathewson77}) observationally mapped the Magellanic Stream and discovered its clumpy structure consisting
of six major H\,I clouds named MS\,I--VI. Recently, a more sensitive high--resolution H\,I survey of the Magellanic System by
Br\"{u}ns et al.~(\cite{Bruens05}) showed that the above mentioned fragments of the Stream have to be considered density peaks
of an otherwise smooth distribution of neutral hydrogen with a column density decreasing towards the high--negative radial velocity tip
of the Magellanic Stream.
Our models corresponding to aspherical MW halos
(A, C) placed local density maxima of H\,I close to the South Galactic pole. That result is supported by observations by Br\"uns
et al.~(\cite{Bruens05}). In this respect, model B did not succeed, and its projected distribution of H\,I in a trailing tail cannot be
considered a satisfactory fit of the Magellanic Stream.

Our models overestimate the integrated relative column densities of H\,I in the part of the
Magellanic Stream located between the South Galactic pole and the far tip of the Stream.
There is also no indication of the H\,I
density decrease as we follow the Stream further from the Magellanic Clouds. In general, all of the models A, B, and C predict
the trailing tail to be of higher H\,I column densities and extended well beyond the far tip of the Magellanic Stream.
Such behavior is a common feature of pure tidal evolutionary models of the Magellanic System and it is a known drawback
of omitting dissipative properties of the gaseous medium.

Regarding the LSR radial velocity measurements along the Magellanic Stream by Br\"uns et al.~(\cite{Bruens05}), our models were able 
to reproduce some of their results. The far tip of the Magellanic Stream in 
model A reaches the extreme negative LSR radial velocity of $\mathrm{-400\,km\,s^{-1}}$ known from H\,I observations.
However, the highest negative LSR radial velocity does not drop below $\mathrm{-350\,km\,s^{-1}}$ for either prolate or nearly spherical halo
configurations. Our previous discussion of various models of the Magellanic System denoted that
the successful reproduction of the high--negative LSR radial velocity at the far tip of the Magellanic Stream is one of the most
challenging problems for such studies. Regarding our results, the importance of the correct LSR radial velocity profile along the Magellanic
Stream was emphasized again. The absence of H\,I between LSR radial velocities of $\mathrm{\approx -350\,km\,s^{-1}}$ and
$\mathrm{\approx -400\,km\,s^{-1}}$ turned out to be the crucial factor decreasing the resulting fitness of examined
evolutionary scenarios.
Based on the study of debris of the Sagittarius dwarf
galaxy, Law et al.~(\cite{Law05}) conclude that the velocity gradient along
the trailing stream provides a sensitive evidence for the mass of the
MW within 50\,kpc and for the oblate shape of the galactic halo. 
However, the leading debris provide a contradictory evidence in favor 
of its prolate shape.

\subsection{Leading Arm}
Reproduction of the Leading Arm remains a difficult task for all the models of the Magellanic System that have been employed so far.
Tidal models naturally place matter to the leading side of the System, towards the Galactic equator, as a result of the tidal stripping
also forming the trailing tail. However, neither the projected shape of the modeled leading structures nor the H\,I density distribution
in the regions having an observational counterpart can be considered sufficient (see, e.g., Gardiner et al.~\cite{Gardiner94}).

In every case A, B, and C, we were able to transport matter to the area of the Leading Arm. Nevertheless, the projected coverage
of that region was more extended than what is observed. All the models contain a significant content of matter spread from the leading edge of LMC
across the Galactic equator, which has not been confirmed observationally. The model C reproduced the Leading Arm
best. But the column density values of H\,I in C are overestimated
and we also could not avoid an additional low--density envelope surrounding the structure (Fig.~\ref{pic_mod_c1_dens}).

\section{Uncertainties in our modeling}\label{missing_stuff}
\subsection{Missing physics}
To optimize performance of the GA,
a computationally fast model of the Magellanic System is required.
Therefore, complex N--body schemes involving self--consistent descriptions of
gravity and hydrodynamics
(see Bekki\,\&\,Chiba~\cite{Bekki05}, Mastropietro et al.~\cite{Mastropietro05})
are discriminated. On the other hand, a correct
description of physical processes dominating the evolution of
the System remains a crucial constraint on the model.

In Sect.~\ref{models} we discussed the applicability of restricted N--body schemes
on problems of galactic encounters and showed that they allow for modeling
of extended streams and tails.
Thus, we devised a restricted N--body code based on the
numerical models by Murai\,\&\,Fujimoto~(\cite{Murai80}) and Gardiner
et al.~(\cite{Gardiner94}). 
The test--particle code interprets the observed
large--scale structures such as the Magellanic Stream or the Leading
Arm as products of tidal stripping in the Magellanic System.

In addition to tidal schemes, ram pressure models have also been used
in the previous studies on the Magellanic System (see Sect.~\ref{models}).
Compared to tidal schemes, hydrodynamical models allow for better reproduction
of the H\,I column density in the Magellanic Stream, which is significantly overestimated
if pure tidal stripping is assumed. Nevertheless, ram pressure models face difficulties
concerning the clumpy structure of the Stream (see Br\"uns et al.~\cite{Bruens05}), which can hardly be
reproduced by the process of continuous ram pressure stripping of the LMC gas as the Cloud moves through
the halo of MW. On the other hand, the results of our study indicate that tidal stripping may produce
inhomogeneous distribution of H\,I in the Magellanic Stream (see Fig.~\ref{pic_mod_a1_dens}) due to the time variations of
the tidal force exerted on the Clouds. Moreover, it is a significant drawback of ram pressure models that they
constantly fail to reproduce the observed slope of the LSR radial velocity profile along the Magellanic Stream
and especially the high--negative velocity tip of the Stream (Fig.~\ref{pic_obs_radvel}).
Generally speaking, both classes of numerical schemes (tidal versus ram pressure) have introduced
promising results into the research of the Magellanic System. Surprisingly, a study combining
the effects of the tidal stripping due to the LMC--SMC--MW encounters with the influence of ram pressure
stripping is still missing. The recent papers
either do not focus on the large scale structure of the Magellanic System
(e.g., Bekki\,\&\,Chiba~\cite{Bekki05}), or ommit the role of gravitational encounters
by neglecting the LMC--SMC interaction and restricting the studied evolutionary period of the System (see, e.g.,
Mastropietro et al.~\cite{Mastropietro05}).

Both families of models suffer from serious difficulties when modeling the Leading Arm.
The tidal models are able to place matter into the area of the Leading Arm, since the creation
of a trailing tail (the Magellanic Stream) is naturally accompanied by the evolution of a leading stream (for
details see Toomre\,\&\,Toomre~\cite{Toomre72}). However, neither the projected shape of the modeled Leading Arm
nor the local column density of H\,I satisfy the observations. Implementation of a hydrodynamical scheme may
improve the results concerning the spatial extent of the tidal Leading Arm and its H\,I column density profile,
due to ram pressure exerted by the MW gas on the leading side of the Clouds. Thus, it is another argument for further
consideration to be given to a hybrid tidal+ram pressure model of the LMC--SMC--MW interaction.

From a technical point of view, employing even a simple formula for ram pressure stripping
would introduce other parameters including structural parameters of
the distribution of gas in the MW halo and a description of the gaseous
clouds in the LMC and SMC. It would increase the dimension of the
parameter space of the interaction and complicate the entire GA
optimization process.

\subsection{Mass and shape evolution of the Magellanic Clouds}
The dark matter halo of the MW is considered axisymmetric and
generally flattened in our model. It is a significant improvement over
previous studies of the Magellanic System that assumed spherical halos
only. We were able to investigate the influence of the potential
flattening parameter $q$ on the evolution of the Magellanic System.
However, both the mass and shape of the MW DM halo were fixed for the
entire evolutionary period of 4\,Gyr.

We did not take into account possible changes in mass and
shape of the Clouds. Shape modification might become important for
very close LMC--SMC encounters that are typical for the models with
nearly--spherical MW DM halos. Pe\~narrubia et al.~(\cite{Pena04})
demonstrated that a relative mass--loss of a satellite galaxy moving
through an extended halo strongly differs for various combination of
its orbital parameters, shape, and the mass distribution of the halo, and
cannot be described reliably by a simple analytic formula.

\section{Summary and conclusions}
We performed an extended analysis of the parameter space for the interaction of the Magellanic
System with the Milky Way. The varied parameters cover the phase space parameters,
the masses, the structure, and the orientation of both Magellanic Clouds as well as the
flattening of the dark matter halo of the MW. The analysis was done by
a specially adopted optimization code searching for a best match between
numerical models and the detailed HI map of the Magellanic System by Br\"uns et al.~(\cite{Bruens05}).
The applied search algorithm is a genetic algorithm combined with a code based on
the fast, but approximative restricted N--body method. By this, we were able to 
analyze more than $\mathrm{10^6}$ models, which makes this study one of the most extended ones
for the Magellanic System.

In this work we focused especially on the flattening of the MW dark matter halo potential $q$.
The range $0.74 \leq q \leq 1.20$
was studied. It is equivalent to the interval of the density flattening $0.31 \leq q_\rho \leq 1.37$
(see Eq.~\ref{log_dens_flattening_2}).

We showed that the creation of a trailing tail (Magellanic Stream) and a leading stream (Leading Arm) is
quite a common feature of the LMC--SMC--MW interaction, and such structures were modeled across the entire
range of halo flattening values.
However, important differences exist between the models, concerning
density distribution and kinematics of H\,I, and also dynamical evolution of the Magellanic System
over the last 4\,Gyr.
In contrast to Murai\,\&\,Fujimoto~(\cite{Murai80}), Gardiner et al.~(\cite{Gardiner94}), or Lin et al.~(\cite{Lin95}),
the Clouds do not have to be gravitationally bound to each other for the entire evolutionary period
to produce the matter distribution that is in 
agreement with currently available H\,I data on the Magellanic System.

Overall agreement between the modeled and observed distribution of neutral hydrogen in the System is quantified by
the fitness of the models. The fitness value is returned by a $FF$, that performs a very detailed evaluation
of every model (Appendix~\ref{appendixB}).
Analysis of fitness as a function of the halo flattening parameter $q$ indicates that the models assuming
oblate DM halo of MW (model A) allow for better satisfaction of H\,I observations than models with other
halo configurations.
Analysis of Fig.~\ref{pic_fitness-q_max} does not indicate a drop in the value of fitness
as the flattening parameter $q$ decreases. It suggests that models of a quality comparable to the fitness of model A
may exist for even more oblate configurations of the MW DM halo. Unfortunately, the logarithmic potential
defined by Eq.~\ref{log_gravity} cannot be used for $q<1/\sqrt{2}$ due to negative density on the axis of symmetry.
Thus, the result represented by Fig.~\ref{pic_fitness-q_max} established a strong motivation for further extension
of the parameter study using a more general and realistic model of the Galactic halo (e.g., a tri--axial
distribution of matter or the flattening parameter depending on position).
Nevertheless, the above mentioned problem does not alter our result preferring flattened halo configurations
and discriminating nearly--spherical shapes within the galactocentric radius of $\approx 200$\,kpc containing
typical orbits of the Clouds and so analyzed in this study.

We did not involve surveys of stellar populations in the Magellanic System in the process of fitness calculations.
This is due to the nature of test--particle models that do not allow for the distinction
between stellar and gaseous content of studied systems. However, we still have to face one of the most interesting
observational facts connected to the Magellanic Clouds -- the absence of stars in the Magellanic Stream
(Van den Bergh~\cite{Vandenbergh00}) -- because both the LMC and SMC contain stellar populations,
and so every structure emanating from the Clouds should be contaminated by stars. It is an additional
constraint on the models. It cannot be involved in fitness calculation because of the limits
of our numerical code, but has to be taken into account.

Stellar populations of the SMC are very young and the mass fraction in the form of stars is extremely low. 
Our models show that the evolution of the Magellanic Stream has been lasting $\mathrm{2\,Gyr}$ at least (model B).
Thus, the fraction of matter in the Magellanic Stream, that is of SMC origin, was torn off
before significant star formation bursts occurred in the SMC, and stars should not be expected in the Stream.
Nevertheless, we found both LMC and SMC matter in the Magellanic Stream for every model of the System.
Similarly to the case of the SMC, if the LMC star formation activity
was increased after the matter transport into the Magellanic Stream was triggered, stars would naturally be missing
in the Stream. Such a scenario is doubtful however. Observational studies 
argue for a massive star formation burst started in LMC at $\mathrm{\approx -3\,Gyr}$
(Van den Bergh~\cite{Vandenbergh00}), which is rather close to the age of the Magellanic
Stream, as indicated by our models (see Sect.~\ref{stream_struct}).
Our results concerning the LMC and SMC orbits introduced an acceptable solution to the
problem of missing stars.
We showed that the evolution of the Clouds in aspherical MW DM halos (models A and C)
does not lead to extremely close encounters
disturbing inner parts of the LMC disk ($r_\mathrm{disk} < 5$\,kpc).
Since the distribution of gaseous matter in galaxies is typically more extended than the stellar content,
the Magellanic Stream matter coming
from outer regions of the Clouds does not necessarily have to contain a stellar fraction.

Previous discussion of stellar content of the Magellanic System supports discrimination
of the configurations with nearly spherical halos (model B) that was discovered by the GA search.
On the other hand,
many papers on the dynamical evolution of the Magellanic Clouds dealing with a spherical MW halo
(Murai\,\&\,Fujimoto~\cite{Murai80},
Gardiner et al.~\cite{Gardiner94}, Bekki\,\&\,Chiba~\cite{Bekki05}) argue that the 
observed massive LMC star formation bursts $\mathrm{3\,Gyr}$ ago was
caused by close LMC--SMC encounters. 
Our model B shows close approaches of 
the Clouds $\Delta r \approx 10$\,kpc at around the mentioned time.
For aspherical halos, such encounters do not induce the formation of 
particle streams. However, close LMC--MW and SMC--MW encounters appeared to be efficient enough to trigger 
massive matter redistribution in the System leading to formation of the observed structures. Then, they could also be
responsible for the triggering of star bursts.

\begin{acknowledgements}
The authors gratefully acknowledge support by the Czech--Austrian
cooperation scheme AKTION (funded by the Austrian Academic Exchange
Service \"OAD and by the program Kontakt of the  Ministery of
Education of the Czech Republic) under grant A--13/2005,
by the Institutional Research Plan AV0Z10030501
of the Academy of Sciences of the Czech Republic and by the project
LC06014 Center for Theoretical Astrophysics.
We also thank Christian Br\"uns who
kindly provided excellent observational data, and Matthew Wall for his unique C++ library for building
reliable genetic algorithm schemes.
\end{acknowledgements}


\clearpage
\appendix

\section{Dynamical friction}\label{appendixA}

If the distribution function in velocity space is axisymmetric, the zeroth order specific friction force is
(Binney~\cite{Binney77}):
\begin{equation}
F_\mathrm{DF}^i = -\frac{2\sqrt{2\mathrm{\pi}}\rho_\mathrm{L}(R,z)G^2M_\mathrm{S}\sqrt{1 - {e_\upsilon}^2}\ln{\Lambda}}{{\sigma_R}^2\sigma_z}B_R\upsilon_i,
\label{dyn_fric_xy}
\end{equation}
\begin{equation}
F_\mathrm{DF}^z = -\frac{2\sqrt{2\mathrm{\pi}}\rho_\mathrm{L}(R,z)G^2M_\mathrm{S}\sqrt{1 - {e_\upsilon}^2}\ln{\Lambda}}{{\sigma_R}^2\sigma_z}B_z\upsilon_z,
\label{dyn_fric_z}
\end{equation}
where $i = x,y$ and $(\sigma_R, \sigma_z)$ is the velocity dispersion ellipsoid with ellipticity
${e_\upsilon}^2 = 1 - (\sigma_z/\sigma_R)^2$, $\mathrm{ln{\Lambda}}$ is the Coulomb logarithm
(Chandrasekhar~\cite{Chandra43}) of the halo, $M_\mathrm{S}$ is the satellite mass, and
\begin{equation}
B_R = \int\limits _0^\infty\frac{\exp{\left(-\frac{\upsilon_R^2/2\sigma_R^2}{1 + q} - \frac{\upsilon_z^2/2\sigma_R^2}{1 -
e_\upsilon^2 + q}\right)}}{(1 + q)^2(1 - e_\upsilon^2 + q)^{1/2}}dq,
\label{dyn_fric_koef_BR}
\end{equation}
\begin{equation}
B_z = \int\limits _0^\infty\frac{\exp{\left(-\frac{\upsilon_R^2/2\sigma_R^2}{1 + q} - \frac{\upsilon_z^2/2\sigma_R^2}{1 -
e_\upsilon^2 + q}\right)}}{(1 + q)(1 - e_\upsilon^2 + q)^{3/2}}dq,
\label{dyn_fric_koef_Bz}
\end{equation}
where $(\upsilon_R, \upsilon_z)$ are the components of the satellite velocity in cylindrical coordinates.

\section{Fitness function}\label{appendixB}
The behavior of the 3\,D test--particle model of the Magellanic System
is determined by a large set of initial conditions and parameters that
can be viewed as a point (individual) in the system's
high--dimensional parameter space.  In the case of our task, the
fitness of an individual means the ability of the numerical model to
reproduce the observed H\,I distribution in the Magellanic Clouds if
the individual serves as the input parameter set for the model. It is
well known that proper choice for $FF$ is critical for the efficiency
of GA and its convergence rate to quality solutions. After extended
testing, we devised a three--component $FF$ scheme. To discover
possible unwanted dependence of our GA on the specific choice for the
$FF$, both of the following $FF$ definitions were employed:
\begin{equation}
FF_a = FF_1 \cdot FF_2 \cdot FF_3,
\label{fitnessA}
\end{equation}
\begin{equation}
FF_b = \frac{\sum\limits_{i=1}^{i=3}{c_i \cdot FF_i}}{\sum\limits_{i=1}^{i=3}{c_i}},
\label{fitnessB}
\end{equation}
where the components $FF_1$, $FF_2$, and
$FF_3$ reflect significant features of the observational data
and $c_1 = 1.0$, $c_2 = 4.0$, and $c_3 = 4.0$
are weight factors.  Both the FFs return values from the
interval $\langle 0.0, 1.0 \rangle$.

FF compares observational data with its models. To do that,
resulting particle distribution has to be treated as neutral hydrogen
and converted into H\,I emission maps for the defined radial velocity
channels.  In the following paragraphs we briefly introduce both the
observed and modeled data processing.

It was shown by Gardiner et al.~(\cite{Gardiner94}) and
Gardiner\,\&\,Noguchi~(\cite{Gardiner96}) that the overall H\,I
distribution in the Magellanic System (Magellanic Stream, Leading Arm) can be considered a
tidal feature. Following that result, an elaborate scheme of the
original data (Br\"uns et al.~\cite{Bruens05}) manipulation was
devised to emphasize large--scale features of the Magellanic H\,I
distribution on the one hand, and to suppress small--scale structures on the
other hand, since they originate in physical processes missing in our
simple test--particle model. The observational data are stored in the
\begin{figure}[h]
\centering
\includegraphics[bb=85 180 500 760, angle=90, width = 8.8cm, clip]{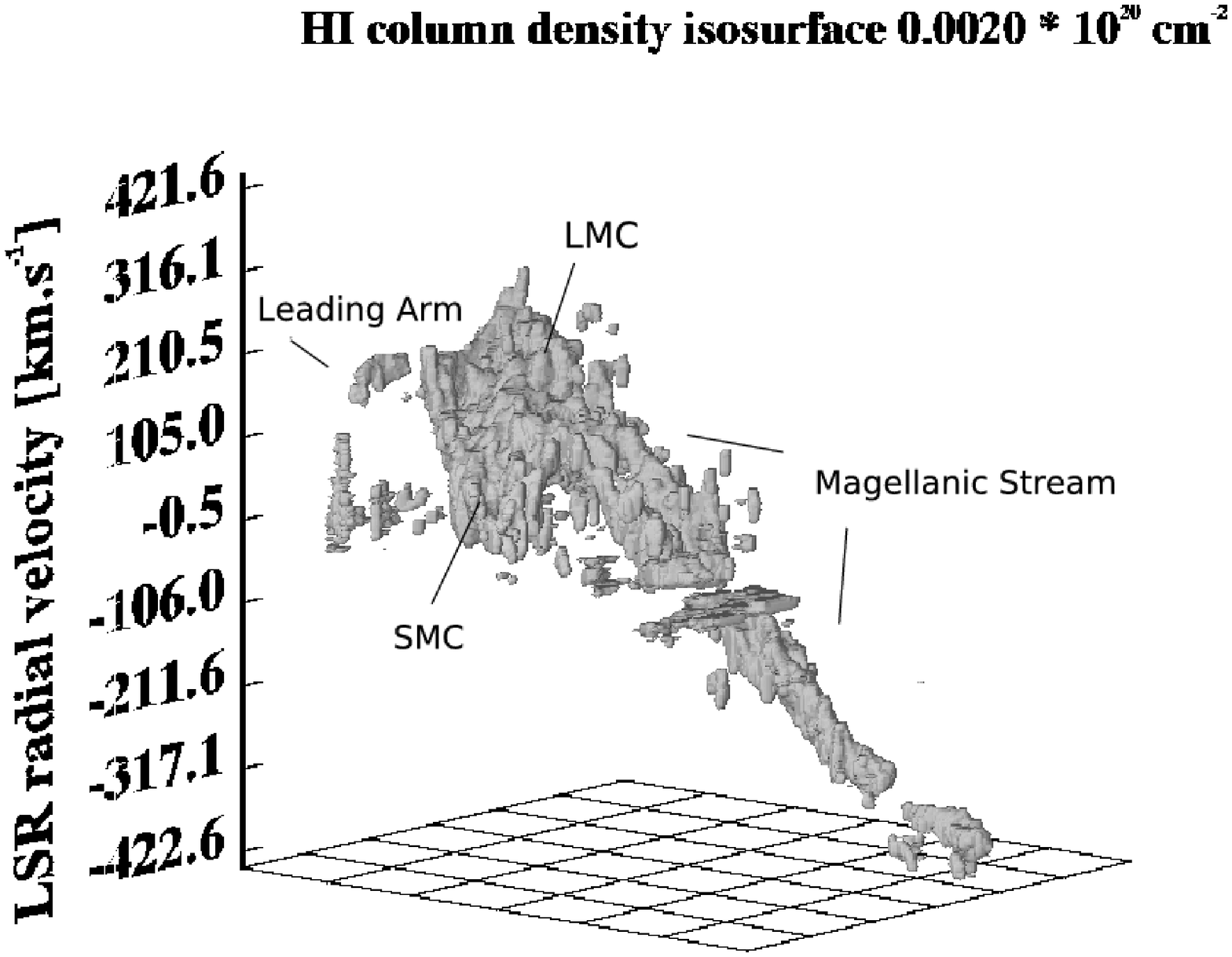}
\includegraphics[bb=110 175 490 700, angle=90, width = 8.8cm, clip]{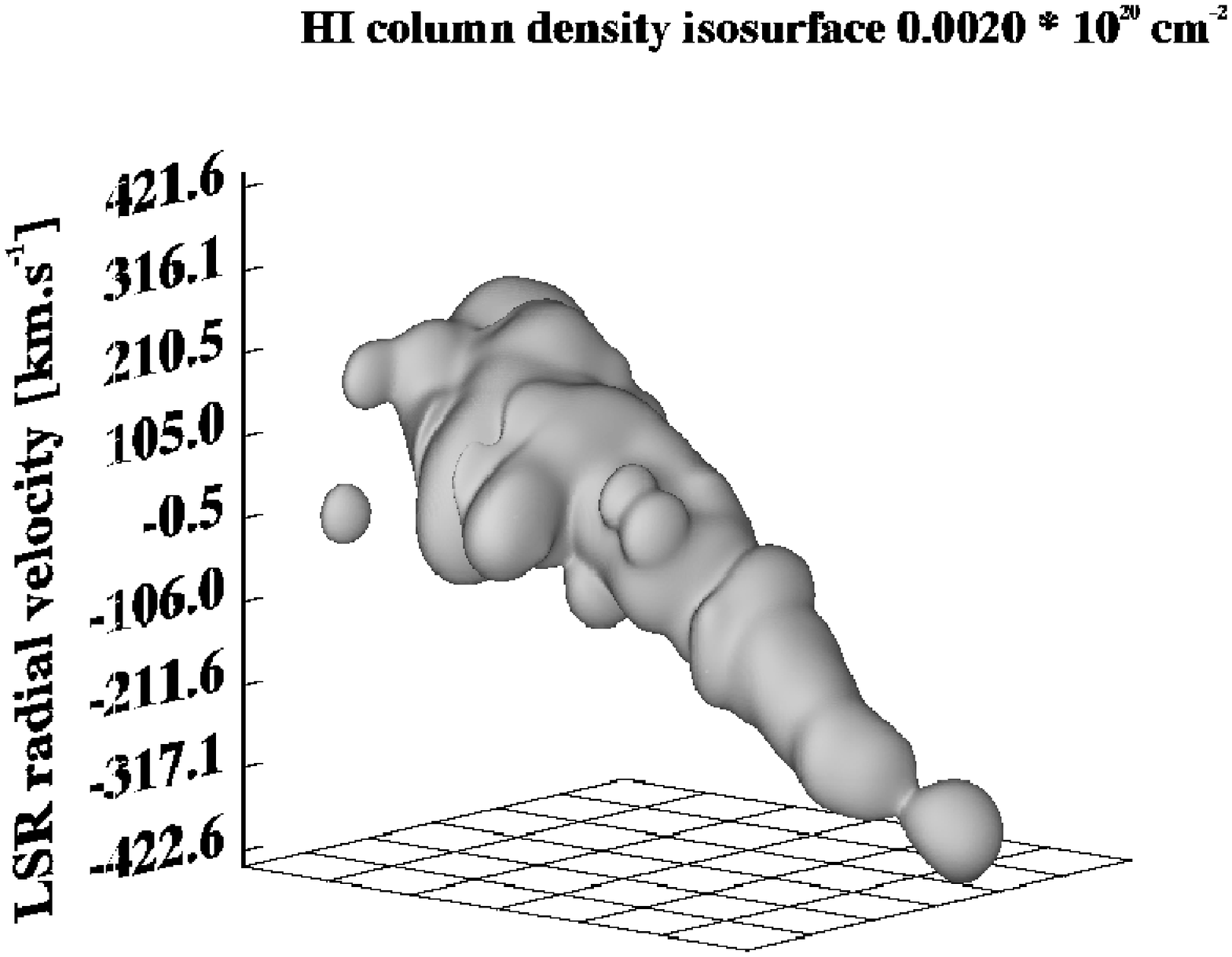}
\caption{The figure depicts the original 3\,D H\,I data cube by Br\"uns et al.~(\cite{Bruens05})
(upper plot) together with the resulting data after median and Fourier filtering. Both images offer
3\,D visualization of the column density isosurface $\mathrm{\Sigma_{H\,I} = 0.2 \cdot 10^{18}\,cm^{-2}}$.}
\label{3d_HI_map}
\end{figure}
Flexible Image Transport System (FITS) format, which lets us apply
standard
image processing methods naturally. A Fourier filter was selected for
our task. It represents a frequency domain filter, and so it allows
for an excellent control over the scale range of the image's
structures to be conserved or filtered out. We removed the wavelengths
below the limit of $\mathrm{\approx 10^\circ}$ projected on the
sky--plane. The performance of Fourier filters suffers from the presence of abrupt changes
of intensity, such as edges and isolated pixels. To enhance the efficiency of
frequency filtering, it was preceded by an application of a spatial median filter to smear the
original image on small scales.
Subsequently, the H\,I column density is normalized. The
resulting 3\,D H\,I column density data cube together with the original data by Br\"uns et al.~(\cite{Bruens05})
can be seen in
Fig.~\ref{3d_HI_map}. To compare the modeled particle distribution with
H\,I observations,
we convert the distribution to
a 3\,D FITS image of column densities
that are proportional to particle counts, since all the
test--particles have the same weight factor assigned.  Then, we have
to interpolate missing data, which is due to a limited number of
particles in our simulations. Finally, the column density is
normalized to the maximal value.

After discussing the data processing and manipulation, we will
introduce the individual FF components $FF_1$,
$FF_2$, and $FF_3$.

\subsection{$FF_1$}
The observed H\,I LSR radial velocity profile measured along the Magellanic Stream is
a notable feature of the Magellanic System.  It shows a linear
dependence of LSR radial velocity on Magellanic Longitude, and a high
negative velocity of $\mathrm{-400\,km\,s^{-1}}$ is reached at the Magellanic Stream's
far tip (Br\"uns et al.~\cite{Bruens05}).  From the studies by
Murai\,\&\,Fujimoto~(\cite{Murai80}) and Gardiner et
al.~(\cite{Gardiner94}), and from our modeling of various Magellanic
evolutionary scenarios, we know that the linearity of the Magellanic Stream velocity
profile shows low sensitivity to the variation of the initial conditions of
the models. On the other hand, the slope of the LSR radial velocity
function is a very specific feature, especially strongly dependent on
the features of the orbital motion of the Clouds. Therefore, it turned
out to be an efficient approach to test whether our modeled particle
distribution was able to reproduce the high negative LSR redial
velocity tip of the Magelanic Stream. Then, the first FF component $FF_1$
was defined as follows:
\begin{equation}
FF_1 = \frac{1}{1 + \left|\frac{\upsilon_\mathrm{min}^\mathrm{obs} - \upsilon_\mathrm{min}^\mathrm{mod}}{\upsilon_\mathrm{min}^\mathrm{obs}}\right|},
\label{fitness1}
\end{equation}
where $\mathrm{\upsilon_{min}^{obs}}$ and
$\mathrm{\upsilon_{min}^{mod}}$ are the minima of the observed LSR
radial velocity profile of the Magellanic Stream and its model, respectively.

\subsection{$FF_2$ and $FF_3$}
The FF components $FF_2$ and $FF_3$ compare the
observed and modeled H\,I column density distributions in the
Magellanic System for 64 separate LSR radial velocity channels of
width $\Delta\upsilon = 13.2\,\mathrm{km\,s^{-1}}$. For every velocity
channel, H\,I column density values are available for ($64 \cdot 128$)
pixels covering the entire System. The above introduced 3\,D data was
obtained by modification of the original high--resolution H\,I
data--cube by Br\"uns et al.~(\cite{Bruens05}). Since the test--particle
model is not capable of reproducing small--scale features of the
explored system, filtering and reduction of resolution of the original
data were necessary prior to its use for the purpose of our GA search.

The second FF component analyzes whether there is a modeled H\,I
emission present at the positions and LSR radial velocities where it
is observed. Thus, we measure the relative spatial coverage of the
System observed in H\,I emission by the modeled matter distribution
for every LSR radial velocity channel. No attention is paid to
specific H\,I column density values here. We only test whether both
modeled and observed emission is present at the same pixel of the
position--velocity space. It can be expressed as
\begin{equation}
FF_2 = \frac{\sum\limits_{i=1}^{N_{\upsilon}} \sum\limits_{j=1}^{N_y} \sum\limits_{k=1}^{N_x} pix_{ijk}^\mathrm{obs} \cdot pix_{ijk}^\mathrm{mod}}
{MAX\left(\sum\limits_{i=1}^{N_{\upsilon}} \sum\limits_{j=1}^{N_y} \sum\limits_{k=1}^{N_x} pix_{ijk}^\mathrm{obs},
\sum\limits_{i=1}^{N_{\upsilon}} \sum\limits_{j=1}^{N_y} \sum\limits_{k=1}^{N_y} pix_{ijk}^\mathrm{mod}\right)},
\label{fitness2}
\end{equation}
where $pix_{ijk}^\mathrm{obs}\in\{0,1\}$ and $pix_{ijk}^\mathrm{mod}\in\{0,1\}$ indicate whether there is matter 
detected at the position [i, j, k] of the 3\,D data on the observed and modeled Magellanic System, respectively.
$N_{\upsilon} = 64$ is the number of separate LSR radial velocity channels in our data. $(N_x \cdot N_y) 
= (64 \cdot 128)$ is
the total number of positions on the sky--plane for which observed and modeled H\,I column density values are 
available.

This binary comparison between the observed and modeled data introduces a problem of pure noise pixels present in the
observed data cube because they posses the same weight as the other data, despite their typically very low intensity.
However, our treatment of the original high--resolution data by Br\"{u}ns et al.~(\cite{Bruens05}) involves
spatial median filtering. It smears abrupt intensity changes and removes isolated pixels, which handles
the problem of pure--noise data pixels naturally. The subsequent Fourier filtering decreases the data resolution significantly,
and that also strongly suppresses the influence of original noise pixels. 

As the last step we compare the modeled matter density distribution to
the observation. To do that, both modeled and observed H\,I column
density values are scaled relative to their maxima to introduce
dimensionless quantities. Then, we get
\begin{equation}
FF_3 = \frac{1}{N_{\upsilon} \cdot N_x \cdot N_y}\sum\limits_{i=1}^{N_{\upsilon}} \sum\limits_{j=1}^{N_y} \sum\limits_{k=1}^{N_x}
\frac{1}{1 + \left|\sigma_{ijk}^\mathrm{obs} - \sigma_{ijk}^\mathrm{mod}\right|},
\label{fitness3}
\end{equation}
where $\sigma_{ijk}^\mathrm{obs}$, $\sigma_{ijk}^\mathrm{mod}$ are
normalized column densities measured at the position [j, k] of the
i--th velocity channel of the observed and modeled data, respectively.

\end{document}